\documentclass[final,authoryear,5p,times,twocolumn]{elsarticle}
\usepackage{epsfig}
\usepackage{wrapfig}
\usepackage[export]{adjustbox}
\usepackage{amssymb}
\usepackage{lineno}
\usepackage{rotating}
\usepackage{color,soul}
\usepackage{textcomp}
\usepackage[font=scriptsize,labelfont=bf]{caption}

\usepackage[bookmarks = true, bookmarksnumbered = true, pdfpagemode =None, pdfstartview = FitH, pdfpagelayout = SinglePage, colorlinks = true, urlcolor = blue, citecolor = monbleu]{hyperref}

\newcommand{\beq}{\bigskip\begin{equation}}
\newcommand{\eeq}{\bigskip\end{equation}}

\journal{Icarus}

\begin{document}

\begin{frontmatter}



\title{Saturn's Seasonal Variability from Four Decades of Ground-Based Mid-Infrared Observations}


\author[le]{James S.D. Blake}
\author[le]{L.N. Fletcher}
\ead{leigh.fletcher@leicester.ac.uk}
\author[jpl]{G.S. Orton}
\author[aa]{A. Antu\~{n}ano}
\author[le]{M.T. Roman}
\author[yk]{Y. Kasaba}
\author[tf]{T. Fujiyoshi}
\author[le]{H. Melin}
\author[le]{D. Bardet}
\author[jpl]{J.A. Sinclair}
\author[ms]{M. Es-Sayeh}

\address[le]{School of Physics \& Astronomy, University of Leicester, University Road, Leicester, LE1 7RH, UK}
\address[jpl]{Jet Propulsion Laboratory, California Institute of Technology, 4800 Oak Grove Drive, Pasadena, CA, 91109, USA}
\address[aa]{UPV/EHU, Escuela de Ingenier\'{i}a de Bilbao, Fisica Aplicada, Spain}
\address[yk]{Planetary Plasma and Atmospheric Research Center, Graduate School of Science, Tohoku University, Sendai, Miyagi, 980-8578, Japan}
\address[tf]{Subaru Telescope, National Astronomical Observatory of Japan, 650 North A'ohoku Place, Hilo, HI 96720, USA}
\address[ms]{Institut de Physique du Globe de Paris - Universit\'{e} de Paris, 35 rue H\'{e}l\'{e}ne Brion, 75013 Paris}

%\linenumbers

%%%%%%%%%%%%%%%%%%%%%%%%%%%%%%%%%%%%%%%%%%%%%%
%%%%%%%%%%%%%%%%%%%%%%%%%%%%%%%%%%%%%%%%%%%%%% ABSTRACT
%%%%%%%%%%%%%%%%%%%%%%%%%%%%%%%%%%%%%%%%%%%%%%
\begin{abstract}
A multi-decade record of ground-based mid-infrared (7-25 $\mu$m) images of Saturn is used to explore seasonal and non-seasonal variability in thermal emission over more than a Saturnian year (1984-2022).  Thermal emission measured by 3-m and 8-m-class observatories (notably NASA's Infrared Telescope Facility, Subaru, and ESO's Very Large Telescope) compares favourably with synthetic images based on both Cassini-derived temperature records and the predictions of radiative climate models.  We find that 8-m class facilities are capable of resolving thermal contrasts on the scale of Saturn's belts, zones, polar hexagon, and polar cyclones, superimposed onto large-scale seasonal asymmetries.  Seasonal changes in brightness temperatures of $\sim30$ K in the stratosphere and $\sim10$ K in the upper troposphere are observed, as the northern and southern polar stratospheric vortices (NPSV and SPSV) form in spring and dissipate in autumn.  The timings of the first appearance of the warm polar vortices is successfully reproduced by radiative climate models, confirming them to be radiative phenomena, albeit entrained within sharp boundaries influenced by dynamics.  Axisymmetric thermal bands (4-5 per hemisphere) display temperature gradients that are strongly correlated with Saturn's zonal winds, indicating winds that decay in strength with altitude from the cloud-tops to the $\sim1$-mbar level, and implying meridional circulation cells in Saturn's upper troposphere and stratosphere forming the system of cool zones and warm belts.  Saturn's thermal structure is largely repeatable from year to year (via comparison of infrared images in 1989 and 2018), with the exception of low-latitudes.  Here we find evidence of inter-annual variations because the equatorial banding at 7.9 $\mu$m is inconsistent with a $\sim15$-year period for Saturn's equatorial stratospheric oscillation, i.e., it is not strictly semi-annual.  Either the oscillation has a longer period closer to $\sim20$ years, or its progression is naturally variable and interrupted by tropospheric meteorology (e.g., storms).  Finally, observations between 2017-2022 extend the legacy of the Cassini mission, revealing the continued warming of the NPSV during northern summer in line with predictions of radiative climate models.

\end{abstract}
%%%%%%%%%%%%%%%%%%%%%%%%%%%%%%%%%%%%%%%%%%%%%%
%%%%%%%%%%%%%%%%%%%%%%%%%%%%%%%%%%%%%%%%%%%%%%
%%%%%%%%%%%%%%%%%%%%%%%%%%%%%%%%%%%%%%%%%%%%%%

\end{frontmatter}

%\linenumbers

%%%%%%%%%%%%%%%%%%%%%%%%%%%%%%%%%%%%%%%%%%%%%%
%%%%%%%%%%%%%%%%%%%%%%%%%%%%%%%%%%%%%%%%%%%%%% INTRODUCTION
%%%%%%%%%%%%%%%%%%%%%%%%%%%%%%%%%%%%%%%%%%%%%%
\section{Introduction}
\label{intro}

Saturn's axial tilt of $27^\circ$ subjects its atmosphere to seasonal cycles of temperatures, aerosols, and chemistry during its 29.5-year orbit of the Sun.  Superimposed onto the seasonal hemispheric asymmetries in temperature and composition are smaller-scale phenomena associated with dynamics, such as belt/zone contrasts \citep{07fletcher_temp}, the formation of polar vortices \citep{05orton, 18fletcher_poles}, equatorial stratospheric oscillations \citep{08fouchet, 08orton_qxo}, inter-hemispheric circulation patterns \citep{09guerlet, 12friedson, 22bardet}, and perturbations from giant storms \citep{12fletcher}.  The Cassini spacecraft, which orbited Saturn from July 2004 (planetocentric solar longitude $L_s=293^\circ$, northern winter) to September 2017 ($L_s=93^\circ$, northern summer), provided an unprecedented record of seasonal variability over almost half a Saturn year from solstice to solstice, and almost from perihelion ($L_s=280^\circ$, June 2003) to aphelion ($L_s=100^\circ$, April 2018), as reviewed by \citet{18fletcher_book}.

However, to understand Saturn's atmosphere over longer periods of time, we rely on the extensive time series provided by ground-based observatories before, during, and after the Cassini epoch.  Over the past four decades, Saturn's thermal emission has been observed from numerous facilities, most notably NASA's Infrared Telescope Facility (IRTF), ESO's Very Large Telescope (VLT), and the Subaru Observatory \citep[the former since 1984, the latter two starting in 2005,][]{09fletcher_imaging}.  Observations during the Cassini mission provided mission support, whereas IRTF observations allow us fill the gap between the Voyager-1 (1980, $L_s=8.6^\circ$) and Voyager-2 (1981, $L_s=18.2^\circ$) encounters \citep{81hanel, 82hanel} and Cassini's arrival (2004, $L_s=293^\circ$).  

Mid-IR imaging observations have been previously published to provide snapshots of Saturn at certain times:  for example, central-meridian scans at 11.7 $\mu$m (sensing stratospheric ethane), taken at Palomar in 1973 by \citet{75gillett}, first revealed Saturn's bright South Polar Stratospheric Vortex (SPSV) during southern summer, a warm hood extending some $\sim15^\circ$ latitude from the pole \citep[also observed in 1977-78 in 7.8- and 12.7-$\mu$m bolometric scans from the Mauna Kea Observatory,][]{80sinton}.  \citet{89gezari} then revealed a similarly warm North Polar Stratospheric Vortex (NPSV) in 1989 using 2D mid-infrared imaging; \citet{05orton} discovered a central south polar cyclone (SPC) within the SPSV using Keck in 2004. \citet{08orton_qxo} used images between 1984 and 2007 to explore Saturn's equatorial stratospheric oscillation and its $\sim15$-year period; \citet{11fletcher_storm} and \citet{12fletcher} observed the tropospheric and stratospheric perturbations associated with a giant storm in Saturn's northern hemisphere; and \citet{17fletcher_QPO} investigated how that same storm disrupted the equatorial oscillation.  However, to date there have been few systematic studies of ground-based imaging data over multiple decades to explore interconnected phenomena that evolve in a cyclic fashion.  In particular, the ground-based data presented in this study constrain the seasonal onset, growth, and dissipation of the NPSV and SPSV over more than one Saturn year \citep{89gezari, 05orton, 18fletcher_poles}; as well as the modulation of equatorial stratospheric brightness by Saturn's equatorial stratospheric oscillation \citep{08orton_qxo, 17fletcher_QPO}.

Collectively, these phenomena lead to a complex seasonal progression for Saturn's atmosphere.  This study uses the 1984-2022 archive of mid-IR imaging (Section \ref{obs}) to present a long term seasonal analysis of Saturn's atmosphere, focusing on three themes. Section \ref{cirs_comparison} uses the Cassini record of Saturn's temperatures between 2004-2017 to reproduce the observed emission in ground-based images during southern summer and northern spring.  We demonstrate how observations from 8-m class observatories can be used to retrieve Saturn's thermal structure and the windshear associated with its belts and zones.  Section \ref{extension} then assesses whether the same temperatures, alongside radiative-convective models, can reproduce images from the IRTF in the 1980s \citep{89gezari} to determine the extent of inter-annual variability in Saturn's seasonal response.  Finally, Section \ref{post_cassini} considers how Saturn has changed since the demise of Cassini, using VLT and Subaru data beyond 2017. 

\section{Observations}
\label{obs}

Saturn's mid-IR spectrum is shaped by the collision-induced opacity of hydrogen and helium, absorption bands of tropospheric PH$_3$ and NH$_3$, and emission bands of CH$_4$ and photochemically-produced hydrocarbons (primarily C$_2$H$_2$ and C$_2$H$_6$).  A standard set of imaging filters, as shown in Figure 1 of \citet{09fletcher_imaging}, would include Q-band filters sounding 80-350 mbar tropospheric temperatures from the H$_2$-He continuum between 17 and 25 $\mu$m, and N-band filters sounding 0.5-5 mbar stratospheric temperatures via CH$_4$ emission at 7.9 $\mu$m and C$_2$H$_6$ emission at 12.3 $\mu$m.  Additional N-band filters were sometimes included, sounding tropospheric PH$_3$ (400-800 mbar) between 8-11 $\mu$m, and stratospheric C$_2$H$_2$ (0.5-15 mbar) at 13.7 $\mu$m.  

Reduction of mid-IR imaging follows the scheme outlined by \citet{09fletcher_imaging}.  Saturn is detected differentially on top of the IR-bright sky background via the chopping and nodding technique, subtracting contributions from the sky and telescope.  Image navigation and geometrical registration requires fitting a planetary silhouette over the image to determine latitudes, longitudes, and emission angles, before producing an equirectangular map of the data.  Given the low reliability of absolute calibration via stellar comparison on non-photometric nights, data were calibrated by scaling the observed flux at low latitudes (and excluding ring absorption) to a low-latitude average of Cassini/CIRS data, as described by \citet{09fletcher_imaging} and further detailed below.  Whilst such a calibration is crude, it enables the creation of a consistent dataset for studying relative brightness changes, but would not capture absolute changes in Saturn's global temperatures over time.

\subsection{VLT/VISIR 2008-2022}

Saturn was regularly observed by the Very Large Telescope (VLT) Imager and Spectrometer for the mid-infrared \citep[VISIR,][]{04lagage} between 2008 and 2022.  VISIR was mounted on the 8.2-m UT3 telescope at Paranal for much of this period, albeit with a 3-year period of refurbishment between 2012 and 2015.  Prior to the upgrade, the DRS $256\times256$ BIB detector offered a field of view of 32" with a pixel scale of 0.127"/pixel, producing the selection of images shown in Fig. \ref{figure:1 thermal_images_comics_visir_pt1}.  We include a single February-2004 mosaicked observation from the Keck Long Wave Spectrometer (LWS) as a comparison immediately prior to Cassini's arrival at Saturn \citep{05orton}.  Three wavelengths are shown:  7.8, 12.2 and 17.6 $\mu$m, sounding stratospheric methane, ethane, and tropospheric H$_2$, respectively, although additional images at 10.3, 18.7, and 19.5 $\mu$m were also acquired within a single hour-long observing block.   The diffraction-limited angular resolution ranges from 0.25" at 7.9 $\mu$m to 0.6" at 19.5 $\mu$m, which equates to 800-1900 km for Saturn near opposition, or $0.75-1.8^\circ$ latitude at Saturn's equator.

The spatial sampling of VISIR was substantially improved after 2015, using a new $1024\times1024$ Raytheon Aquarius IBC detector with a pixel scale of 0.045"/pixel over a 38" field of view.  Although Saturn's rings are often `clipped' by this field of view, the excellent plate scale offers the high-quality imaging shown in Fig. \ref{figure:2 thermal_images_comics_visir_pt1}.  Indeed, Saturn's thermal hexagon can be seen in both the troposphere- and stratosphere-sensing filters (particularly between 2016 and 2018, when the north pole was in view), confirming that the sinusoidal variation at the edge of the north-polar vortex extends throughout the troposphere and stratosphere \citep{18fletcher_poles}.

Since the upgrade of the VISIR detector in 2015, high-brightness images have suffered from a pattern of vertical and horizontal stripes that must be removed before use. Two techniques were explored: a Gaussian-smooth of the image to identify and subtract high-frequency stripes, and a Fourier transform of the image to identify periodic structures to mask out, before reconstructing the image in the spatial domain.  Given the strong variation in flux across the VISIR images associated with Saturn's limb and rings, the latter technique was found to be superior, and was used to develop the image montage in Fig. \ref{figure:2 thermal_images_comics_visir_pt1}.  A full list of observation dates and program numbers are provided in Table \ref{datatable}, and raw data are available through the ESO archive\footnote{\url{http://archive.eso.org/eso/eso_archive_main.html}}.

\subsection{Subaru/COMICS 2005-2020}
\begin{figure*}
\begin{centering}
\centerline{\includegraphics[angle=0,scale=.45]{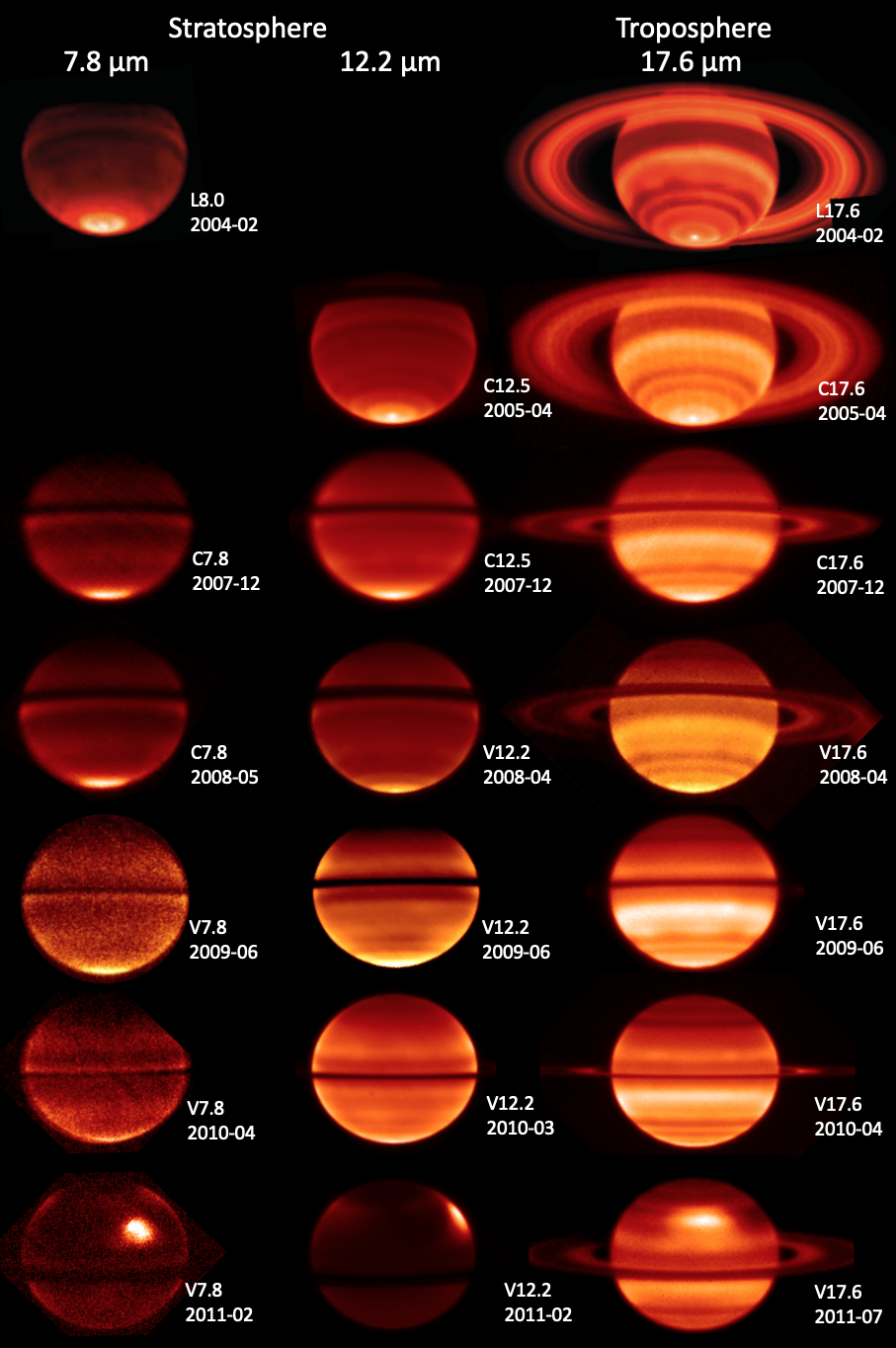}}
\caption{Observations of Saturn from VLT/VISIR (marked with V), Keck/LWS \citep[marked with an L,][]{05orton}, and Subaru/COMICS (marked with C) from 2004 to 2011 for the filters 7.8, 12.2 and 17.65 $\mu$m.  This spans the period before and after the northern spring equinox ($L_s=0^\circ$) in 2009, and 2011 images appear unusual due to the presence of Saturn's northern storm \citep{12fletcher}.}
\label{figure:1 thermal_images_comics_visir_pt1}
\end{centering}
\end{figure*}

\begin{figure*}
\begin{centering}
\centerline{\includegraphics[angle=0,scale=.7]{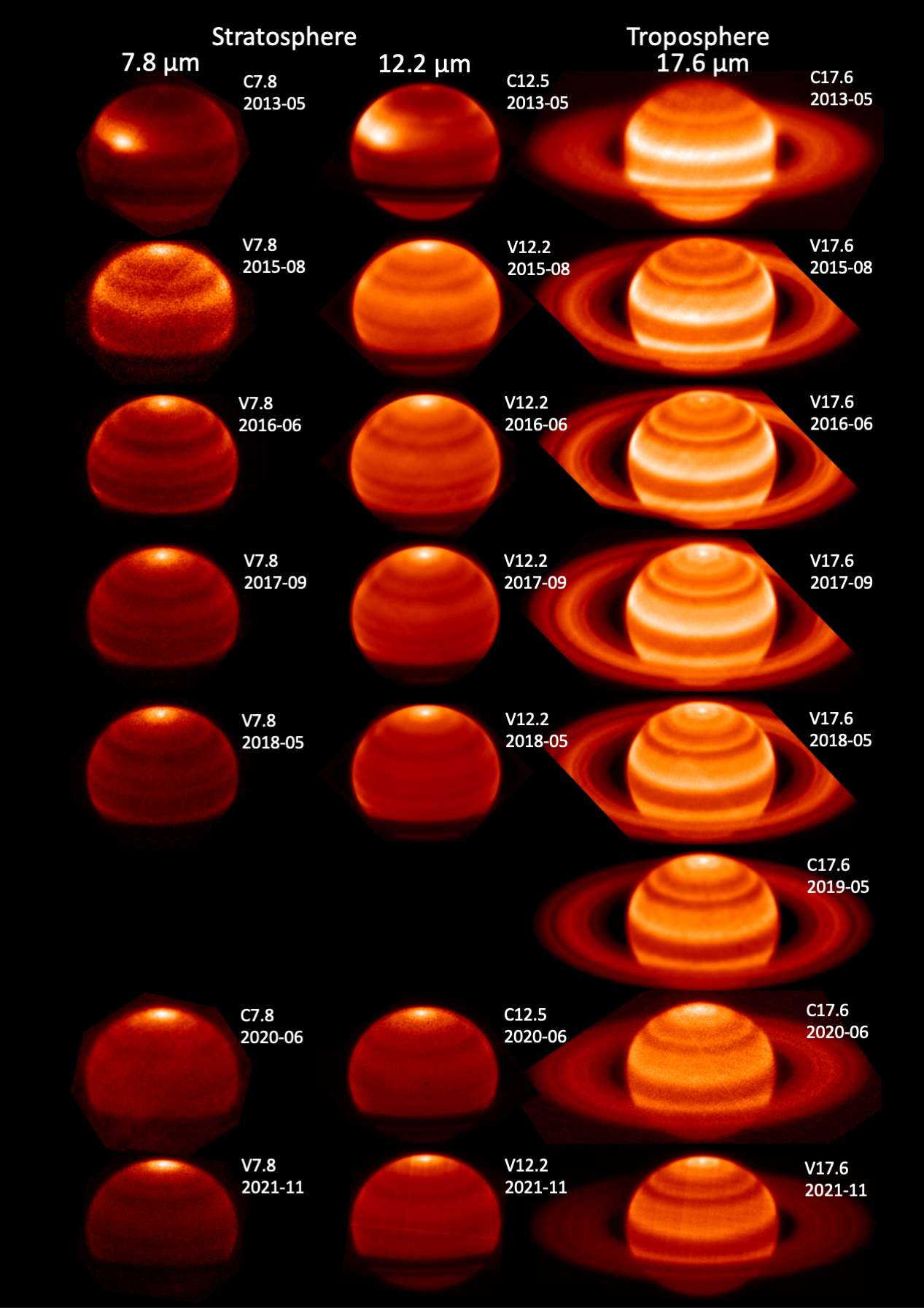}}
\caption{Observations of Saturn from VLT/VISIR (marked with V) and Subaru/COMICS (marked with C) from 2013-2021, spanning the period surrounding Saturn's northern summer solstice ($L_s=90^\circ$) for the filters 7.8, 12.2 and 17.65 $\mu$m.  Note the visibility of Saturn's polar hexagon at 17.6 $\mu$m, surrounding the compact warm polar cyclone (PC). }
\label{figure:2 thermal_images_comics_visir_pt1}
\end{centering}
\end{figure*}

VISIR observations were supplemented by images from the Cooled Mid-IR Camera and Spectrograph (COMICS) on the Subaru telescope of Japan \citep{02kataza, 03okamoto} between 2005 and 2020, a subset of which are shown in Fig. \ref{figure:1 thermal_images_comics_visir_pt1} and \ref{figure:2 thermal_images_comics_visir_pt1}.  The COMICS $320\times240$ array had a coarser plate scale of 0.133"/pixel compared to VISIR, but a larger field of view ($42\times32$"), preventing clipping of Saturn's rings.  Previously-published observations from April 2005 \citep{09fletcher_imaging} spanned 8.7-24.5 $\mu$m, but lacked a 7.8-$\mu$m filter sensitive to stratospheric methane.  This was added for later datasets in each of the following years:  2007-2009; 2013; and 2017-2020.  COMICS was decommissioned in 2020, and raw data are available through the SMOKA archive\footnote{\url{https://smoka.nao.ac.jp/}}.  The datasets used in this study are recorded in Table \ref{datatable}.

\subsection{NASA/IRTF 1984-2003}

Prior to 2003, mid-IR imaging of Saturn was provided primarily by NASA's Infrared Telescope Facility (IRTF).  With its 3-m primary mirror, IRTF provides diffraction-limited imaging with spatial resolutions varying from 0.65-2.0" at 7.9-24.5 $\mu$m, equivalent to 2200-6500 km for Saturn at opposition, or $2-6^\circ$ latitude at the equator.  Additional blurring arises from atmospheric seeing, which varied from night to night.  Between 1984 and 1989, single-element photometers (the BOLO1 and AT1 instruments) were scanned across Saturn to create maps - details of the acquisition and calibration of these data can be found in the supplementary material of \cite{08orton_qxo}. 

\begin{figure*}
\begin{centering}
\centerline{\includegraphics[angle=0,scale=.4]{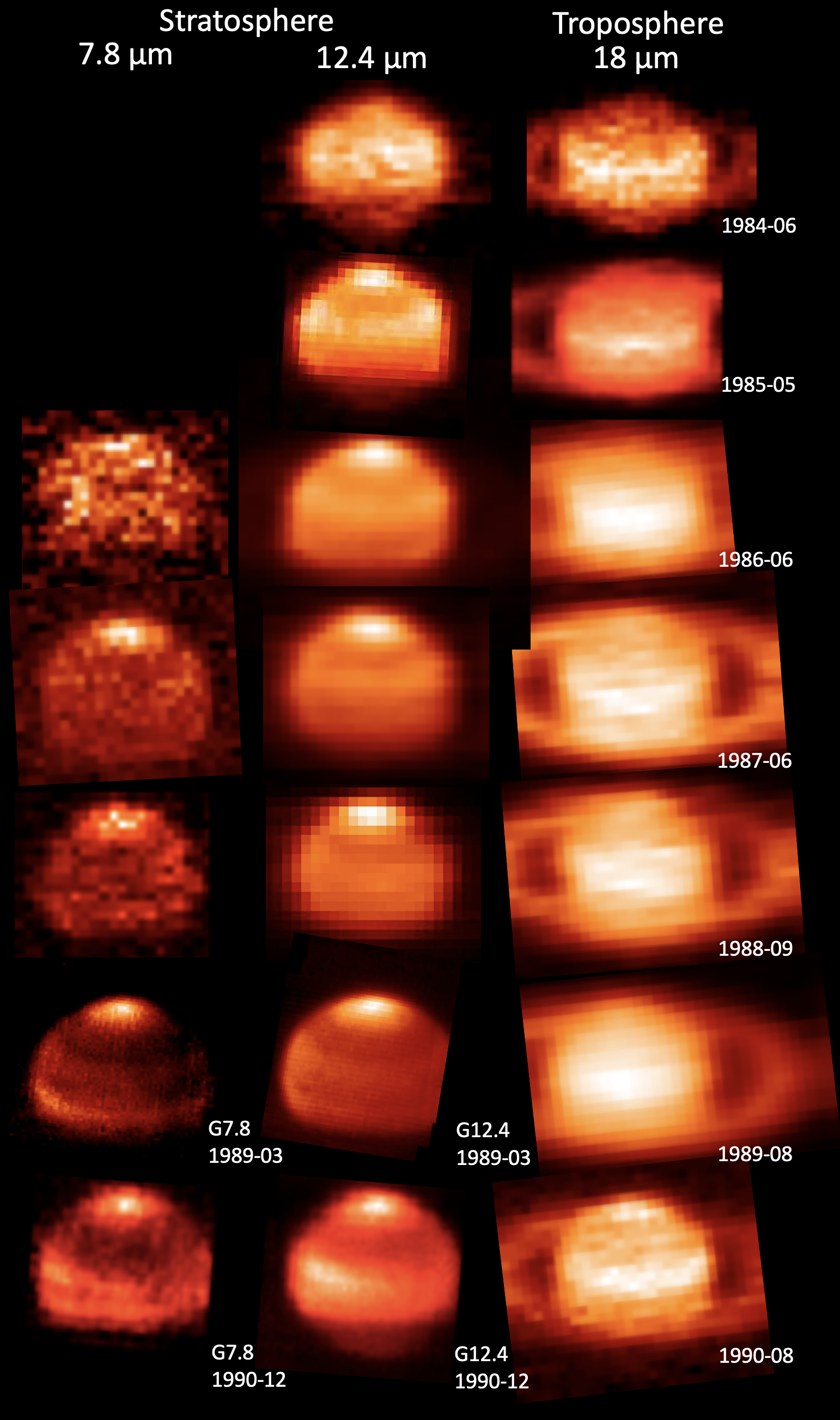}}
\caption{Observations of Saturn between 1984 and 1990 using the IRTF, spanning the period surrounding northern summer solstice in October 1987 ($L_s=90^\circ$). Images are primarily from raster-scans using single-element photometers.  After 1989, observations used the 2D camera of \citet{89gezari}, marked with a `G'.  This montage reveals the development of the warm NPSV, and the December-1990 stratospheric images reveal east-west structure that occurred a few months after the September-1990 Great White Storm \citep{91sanchez}.}
\label{figure:3 thermal_images_GRABER_GEZARI}
\end{centering}
\end{figure*}

Observations by \cite{89gezari} in March 1989 provided the first two-dimensional infrared observations of Saturn from a ground-based telescope (IRTF). The 1989 images were calibrated using the radiance values listed by \citet{89gezari} for the north pole: 2.7, 9.7 and 20.0 Jy arcsec$^{-2}$ for filters at 7.8 $\mu$m, 11.6 $\mu$m, and 12.4 $\mu$m, respectively. The raster-scanned data acquired prior to 1989 were calibrated by scaling it to match the 1989 observations of \citet{89gezari} where possible, as absolute stellar calibration was not available. The 1984-1990 raster-scanned observations of \citet{08orton_qxo} and 1989 images of \citet{89gezari} are shown in Fig. \ref{figure:3 thermal_images_GRABER_GEZARI}, and the available observations have been added to Table \ref{datatable}.  We note that December 1990 images show interesting low-latitude structures, with brighter emission towards the west that was not apparent in the March 1989 images of \citet{89gezari}.  These data were taken approximately three months after the Great White Storm of September 1990 \citep{91sanchez}, and may indicate stratospheric perturbations from the storm, as seen in the more recent storm eruption of 2010-2013 \citep{12fletcher}.

% Additional observations from the 2D camera of \citet{89gezari} are available for December 1990.  

\begin{figure*}
\begin{centering}
\centerline{\includegraphics[angle=0,scale=.7]{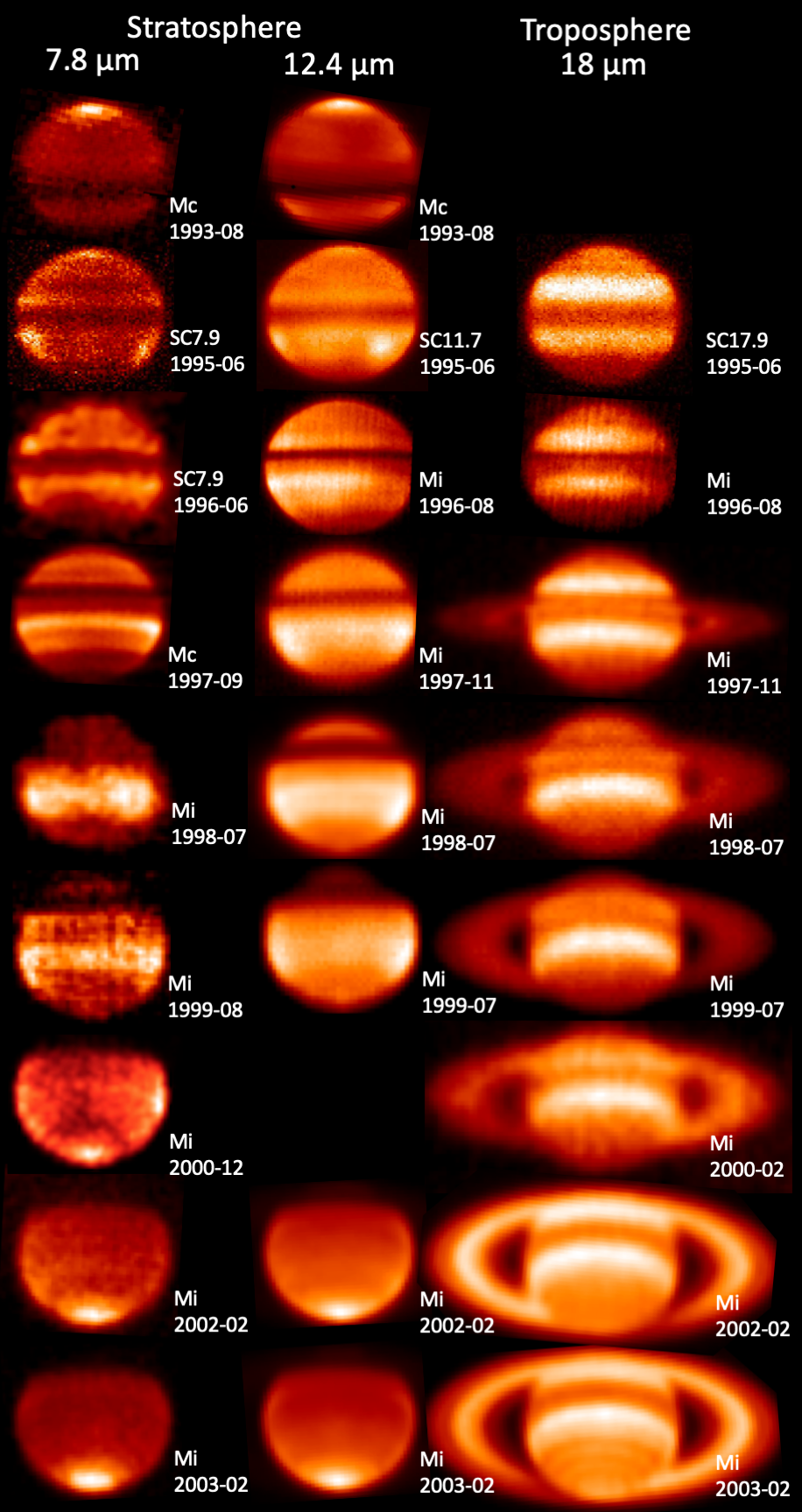}}
\caption{Observations of Saturn from the MIRLIN (marked Mi) and MIRAC2 (marked with Mc) instruments on the IRTF throughout the 1990s, showing the steady improvement in image quality throughout this period.  Observations from SPECTROCAM-10 \citep[mounted on the Hale 200-inch and marked SC,][]{08orton_qxo} are also shown for June 1995 and June 1996, either side of southern spring equinox ($L_s=180^\circ$). }
\label{figure:4 thermal_images_MIRLIN_MIRSI}
\end{centering}
\end{figure*}

From the early 1990s, 2D imaging of Saturn with the IRTF instruments became more frequent.  MIRAC2 (Mid-Infrared Array Camera, 1993-1997) provided imaging with a 0.34"/pixel scale (at 10 $\mu$m) and a maximum field-of-view of $44\times44$" \citep{93hoffmann}.  MIRLIN (1996-2004) was a $128\times128$ pixel, 7-25$\mu$m infrared astronomical camera built at JPL and operated on IRTF, with a pixel scale of 0.475"/pixel and a field of view of $61\times61$" \citep{94ressler} for the same filters as MIRAC2.  MIRLIN and MIRAC2 images for filters at 7.8, 12.4 and 18.0 $\mu$m have been used in the time series in Fig. \ref{figure:4 thermal_images_MIRLIN_MIRSI}, approximately covering the period between southern spring equinox ($L_s=180^\circ$, November 1995) and southern summer solstice ($L_s=270^\circ$, October 2002), and bringing us to the epoch when Cassini observations and 8-m class facilities were observing regularly.  The IRTF images were supplemented in 1995 and 1996 by observations from the SpectroCam-10 instrument on the Hale 200-inch at Palomar \citep[described by][]{08orton_qxo}.

Two additional sources of IRTF data are not considered as part of this study - imaging from the MIRSI instrument between 2003 and 2012 \citep{03deutsch}, and spectral scan maps from the TEXES instrument \citep{02lacy} acquired since 2002 \citep{05greathouse}.  The former are superseded in this study by the aforementioned VISIR and COMICS observations spanning the same time period with a superior spatial resolution. The spectroscopic maps from TEXES require significantly different processing techniques to ensure a valid comparison to the imaging datasets.  These extensive datasets will be the subject of a future study. 

%%%%%%%%%%%%%%%%%%%%%%%%%%%%%%%%%%%%%%%%%%%%%%
%%%%%%%%%%%%%%%%%%%%%%%%%%%%%%%%%%%%%%%%%%%%%% 
%%%%%%%%%%%%%%%%%%%%%%%%%%%%%%%%%%%%%%%%%%%%%%

\section{Comparison of Cassini and Ground-Based Observations 2004-2017}
\label{cirs_comparison}

In this Section, we compare observations from 8-m class observatories (VLT and Subaru) to the record of brightness temperatures and retrieved temperatures from the Cassini/CIRS instrument.  We explore the ground-based data in the following subsections by (i) generating synthetic images using the CIRS database for comparison to observations of Saturn's seasonal changes (Section \ref{synth_images}); (ii) measuring brightness temperature gradients to understand Saturn's belt/zone structure (Section \ref{beltzone}); and (iii) deriving 2D temperature-pressure cross-sections from stacked filtered images (Section \ref{inversion}).

\begin{figure*}
\begin{centering}
\centerline{\includegraphics[angle=0,scale=.4]{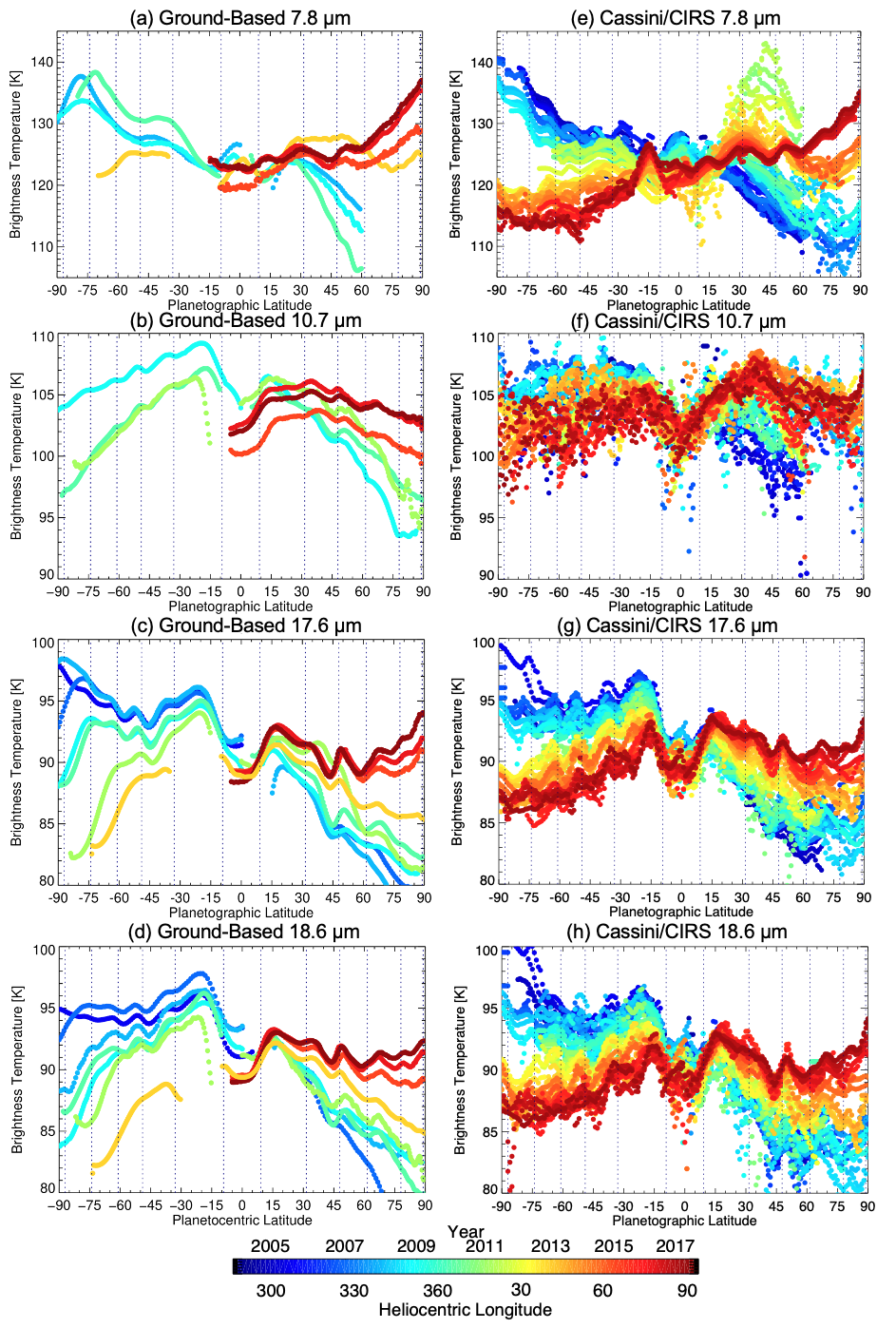}}
\caption{Seasonal changes in Saturn's brightness temperatures between 2004 and 2017, comparing yearly records from VISIR and COMICS (left) to more regular latitudinal scans from Cassini/CIRS (right).  Regions obscured by Saturn's rings are omitted from the ground-based zonal averages.  Measurements are colour-coded by date, according to the shared colour legend at the bottom of the figure.  The location of local maxima in the zonal winds (eastward jets) are marked with vertical dotted lines following \citet{11garcia}.  The stratosphere is revealed by 7.8-$\mu$m observations, whereas 10.7, 17.6 and 18.6 $\mu$m sound the troposphere.}
\label{CIRS-VISIR-Lat-tb-comparison}
\end{centering}
\end{figure*}

Zonal-mean brightness temperatures for COMICS and VISIR observations from 2005-2017 are shown in Fig. \ref{CIRS-VISIR-Lat-tb-comparison}, compared to filter-integrated brightnesses measured by Cassini.  We show a selection of four filters:  7.8, 10.7, 17.6 and 18.6 $\mu$m.  These are not true zonal averages, but approximations formed by averaging within $\pm30^\circ$ longitude of the central meridian for each image, assuming longitudinal homogeneity.  This is a good approximation for most dates except for those in 2011-13, where significant longitudinal structure was present due to the northern springtime storm \citep{12fletcher}, as observed in Fig. \ref{CIRS-VISIR-Lat-tb-comparison}(e).  Regions obscured by Saturn's rings are omitted for each date, although some high-latitude `winter-hemisphere' points are included.  The ground-based zonal-mean $T_B$ is compared to data from Cassini/CIRS - specifically a database of spectra developed by curtailing CIRS interferograms so that they all have the same length (e.g., 15-cm$^{-1}$ spectral resolution) following \citet{17fletcher_QPO}.  The CIRS far-IR focal plane one was used for Q-band observations ($>17$ $\mu$m), whereas the mid-IR focal planes (3 and 4) were used for N-band observations ($7-13$ $\mu$m).  CIRS spectra were convolved with VISIR filter functions to approximate the brightness temperature as seen by the ground-based observatories in Fig. \ref{CIRS-VISIR-Lat-tb-comparison}(e-h).   No attempt has been made to remove limb-brightening/darkening in either dataset.

Fig. \ref{CIRS-VISIR-Lat-tb-comparison} shows the seasonal progression of tropospheric and stratospheric temperatures during the course of the Cassini mission.  In the stratosphere, the southern-summer $T_B$ ranges from 140 K at the south pole to 110 K at the north pole in 2005-2007, but by the end of the Cassini mission in 2017 this gradient had reversed, with peak brightness temperatures of $\sim140$ K at the north pole.  Equatorial stratospheric temperatures at 7.8 $\mu$m show variations associated with Saturn's stratospheric oscillation - from a local maximum in equatorial $T_B$ in 2004-06, to a local minimum in equatorial $T_B$ in 2015-17, consistent with the approximately $\sim15$-year period identified by \citet{08orton_qxo}.  Although the close-match of the absolute brightness temperatures is unsurprising, given our method of radiometric calibration, the fact that the relative changes in $T_B$ are consistent between Cassini and the ground-based record is encouraging.  Seasonal changes in the troposphere are smaller (pole-to-pole contrasts of $\sim10-15$ K), consistent with the longer radiative timescales of the deeper atmosphere \citep{90conrath}.  In almost all filters, 2017 (summer solstice) presented the warmest $T_B$ at the north pole, associated with the North Polar Stratospheric Vortex (NPSV) \citep{18fletcher_poles}.

The latitudes of the eastwards peaks in the cloud-tracked zonal winds \citep{11garcia} are also shown in Fig. \ref{CIRS-VISIR-Lat-tb-comparison}.  This shows qualitatively that $T_B$ contrasts measured by Cassini and VLT/Subaru are co-located with the peak eastward winds.  We assess this quantitatively in Section \ref{beltzone}.

%/data/nemesis/jsdb3/visir/TB_10.7/plot_lat_tb.pro
%/data/nemesis/jsdb3/visir/TB_7.9/plot_lat_tb.pro
%/data/nemesis/jsdb3/visir/TB_17.6/plot_lat_tb.pro
%/data/nemesis/jsdb3/visir/TB_18.7/plot_lat_tb.pro

\subsection{Synthetic Saturn Images based on CIRS Database}
\label{synth_images}

\begin{figure}
\begin{centering}
\centerline{\includegraphics[angle=0,scale=.5]{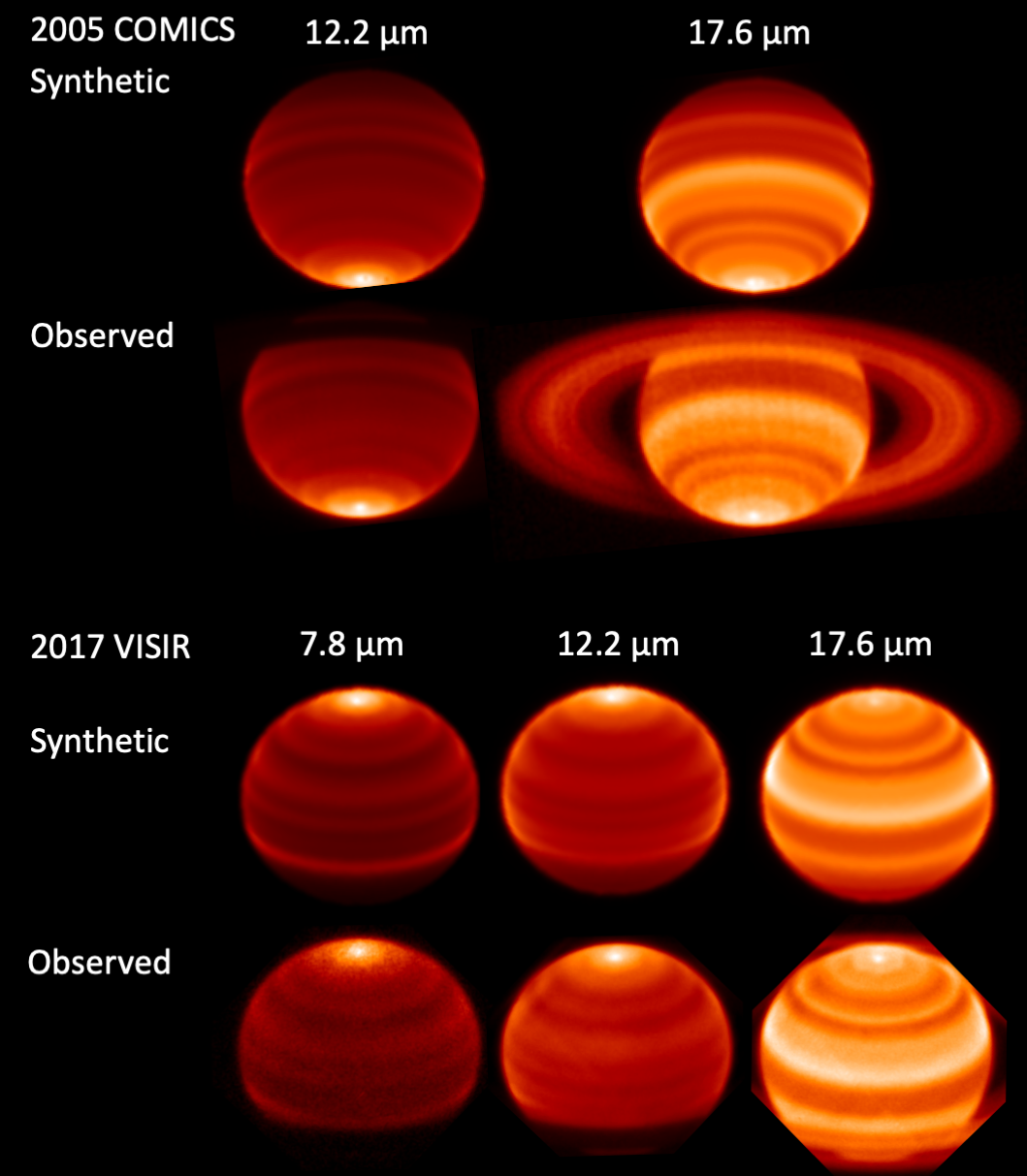}}
\caption{Comparison of VISIR (September 2017) and COMICS (April 2005) observations of Saturn with synthetic images generated using the Cassini-derived record of temperatures and hydrocarbons, as described in the main text.}
\label{synthetic_comparison}
\end{centering}
\end{figure}

The time series of Saturn's atmospheric temperatures and stratospheric composition, retrieved from Cassini/CIRS between 2004 and 2017 \citep[e.g.,][]{16fletcher, 18fletcher_poles}, was used to generate synthetic images of Saturn for comparison with ground-based data in Fig. \ref{synthetic_comparison}.  In this section we focus on COMICS observations in April 2005, and VISIR observations in September 2017.  The latitude, longitude, and emission angle were extracted for every pixel in each image, such that viewing geometry and detector plate scales are properly accounted for.  Latitude-pressure cross-sections of retrieved temperatures, ethane, and acetylene abundances were then used to generate synthetic radiances in each pixel, using the NEMESIS suite of radiative transfer and spectral retrieval tools \citep{08irwin}.  Vertical and latitudinal profiles of CH$_4$, NH$_3$, and PH$_3$ were held fixed, and sources of spectral linedata (used to generate $k$-distributions for rapid calculations of spectral radiance) and collision-induced opacity are as described in \citet{18fletcher_poles}.  The $k$-distributions were convolved with filter profiles and telluric transmissions specific to each instrument, following \citep{09fletcher_imaging}.  

The calculated spectral radiance for each filter was used to assemble a synthetic image, which was then convolved with a wavelength-dependent Airy function (to represent diffraction) and a Gaussian (to represent the seeing).  Noise was added to the synthetic to match that observed in the background surrounding Saturn, and the synthetics are compared to images in Fig. \ref{synthetic_comparison}.  The qualitative comparison is good, showing thermal banding and the increased emission of the SPSV (2005) and NPSV (2017) in the stratosphere, and even the compact polar cyclones (PCs) at 17.6 $\mu$m.  

\begin{figure}
\begin{centering}
\centerline{\includegraphics[angle=0,scale=.8]{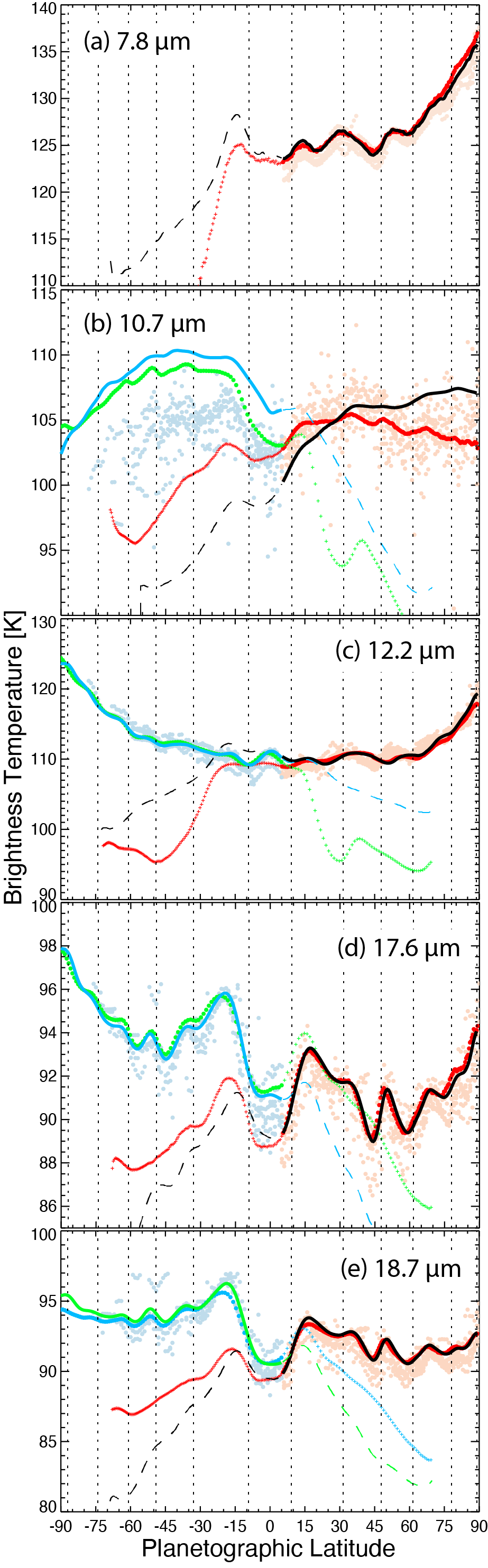}}
\caption{Comparison of VISIR and COMICS brightness temperatures with CIRS-based synthetic images from Fig. \ref{synthetic_comparison}. The September 2017 VISIR data (thick black) is compared with a synthetic profile (thick red) that is derived from the CIRS results.  The dashed thin black line (data) and red crosses (synthetic, appearing as a thin red line) indicate regions obscured by Saturn's rings. Similarly, the 2005 COMICS profile (thick green line) is compared to the synthetic profile (thick blue), with the thinner dashed and crossed lines marking regions of ring obscuration.  CIRS brightness temperatures are shown in the background as scattered points with colours matching the COMICS (2005, light blue) and VISIR (2017, light red) data. Neither COMICS nor VISIR data were available at 7.9 $\mu$m from 2005 for comparison in panel (a). The discrepancy at 10.7 $\mu$m is described in the main text.  The latitudes of eastward maxima in the Cassini-derived wind field \citep{11garcia} are marked as vertical dotted lines. }
\label{synthetic_profile-VISIR-CIRS_comparison}
\end{centering}
\end{figure}

We extract zonal-mean $T_B$ profiles from both the synthetic images and the real data in Fig. \ref{synthetic_profile-VISIR-CIRS_comparison} for a selection of five filters, averaging brightness within $\pm30^\circ$ longitude of the central meridian.  COMICS data in southern summer (blue-green) and VISIR data in northern summer (black-red) show agreement for all filters except 10.7 $\mu$m (see below).  The strong $T_B$ gradients associated with Saturn's belts and zones are co-located in the CIRS-synthetic and the ground-based observations (see Section \ref{beltzone}), except in regions where Saturn's rings begin to obscure the view from Earth.  Polar brightness temperatures in both the troposphere and stratosphere are accurately reproduced by the CIRS synthetic.  Furthermore, the ground-based $T_B$ profiles are largely in the centre of the distribution of the CIRS $T_B$ measurements, shown in the background of Fig. \ref{synthetic_profile-VISIR-CIRS_comparison}.  We conclude from this comparison that the CIRS-derived temperatures and hydrocarbons are consistent with the COMICS and VISIR observations between 2004 and 2017, adding further validation to the CIRS results.

% The latitudinal temperature brightness profiles of the synthetics based on the Cassini CIRS temperature record are largely in strong agreement with the ground-based observations. Visual inspection of the synthetic images in comparison to the VISIR and COMICS observations (Fig. \ref{figure:5 synthetic_comparison}) show a high degree of fidelity in between the two for both instruments. The same latitudinal band structures are clearly discernible in both the synthetics and the observations, with the exception of the rings as these were not modelled. 

% Fig. \ref{figure:6 CIRS-VISIR-Lat-tb-comparison} A-D each show very similar progression in the changes to the latitudinal temperature brightness profiles over time with the CIRS temperature record (figures 9, 10 and 11). Further comparisons for these filters (7.9, 12.3, 17.6 $\mu$m) show that the strong agreement between the observations and the CIRS temperature record based synthetic profiles.

The one exception is the 10.7 $\mu$m filter in Fig. \ref{synthetic_profile-VISIR-CIRS_comparison}(b), where the synthetic is not a good match to the observations.  Although the observations are in a very dark region of Saturn's mid-IR spectrum, Fig. \ref{synthetic_profile-VISIR-CIRS_comparison}(b) nevertheless shows structure.  The 10.7-$\mu$m synthetic has more subdued banding than the data reveal, and the centre-to-limb variation of brightness does not match the observations.  This filter senses the $\sim500$-mbar region of Saturn's troposphere, at depths where the far-IR CIRS focal plane was used to constrain atmospheric temperature.  The downside of this was that the far-IR focal plane had a lower spatial resolution, and did not resolve belt/zone contrasts as well as the shorter-wavelength mid-IR CIRS focal planes \citep{04flasar}.  Furthermore, 10.7 $\mu$m spectral radiance depends on both the temperature structure and the distribution of PH$_3$.  Due to an absence of published time-resolved CIRS PH$_3$ results, PH$_3$ was assumed to be uniform with latitude and time, which could have lessened the contrasts in the 10.7-$\mu$m synthetic. Furthermore, the vertical distribution of PH$_3$ was not allowed to vary, which could cause the mismatch in the centre-to-limb brightness.  The CIRS temperature and hydrocarbon record of \citet{18fletcher_poles} is therefore insufficient to reproduce ground-based observations at 10.7 $\mu$m, and future modelling of Saturn's N-band spectrum (e.g., from the TEXES instrument mentioned in Section \ref{obs}), alongside longer integration times at 10.7 $\mu$m, will be required to improve on this mismatch.

% This is true for all tested filters with the exception of 10.7 $\mu$m emission which shows significant differences both in seasonal progression (Fig. \ref{figure:6 CIRS-VISIR-Lat-tb-comparison}) and the direct comparison (Fig. \ref{figure:7 synthetic_profile-VISIR-CIRS_comparison}). We posit that this may be due to an incomplete understanding of the phosphine emissions in this spectral region, leading to an absence of particular emission features in the line list and hence the ktables used to make the synthetic with NEMESIS. Refinements in the line list through laboratory experimentation would be the primary solution for rectifying this disparity. All other filters show strong agreement but have been omitted for brevity.
 
% From this comparison we can conclude that with the exception of the 10.7 $\mu$m filter, the Cassini/CIRS temperature record is both accurate and capable of replicating ground-based observations through the process of forward modelling with NEMESIS. 

\subsection{Saturn's Belts and Zones}
\label{beltzone}

Observing Saturn with 8-m diameter mirrors provides sufficient spatial resolution to distinguish Saturn's thermal bands in the upper troposphere and stratosphere, as shown in Figs. \ref{figure:1 thermal_images_comics_visir_pt1} and Fig. \ref{figure:2 thermal_images_comics_visir_pt1}.  Latitudinal gradients in atmospheric temperature are related to the vertical shear on Saturn's zonal jets via the geostrophic thermal wind equation \citep[e.g.,][]{04holton}, and the decay of Saturn's zonal jets with altitude has been inferred via spectroscopic inversion using both Cassini \citep{05flasar, 09read, 16fletcher} and Voyager data \citep{81pirraglia}.  In this section, we show that the same conclusion can be reached via correlating brightness temperature contrasts from Subaru/COMICS during southern summer (2005 April 30) and VLT/VISIR during northern summer (2017 September 7).  Note that a full temperature inversion will be performed in Section \ref{inversion}.

\begin{figure*}
\begin{centering}
\centerline{\includegraphics[angle=0,scale=.45]{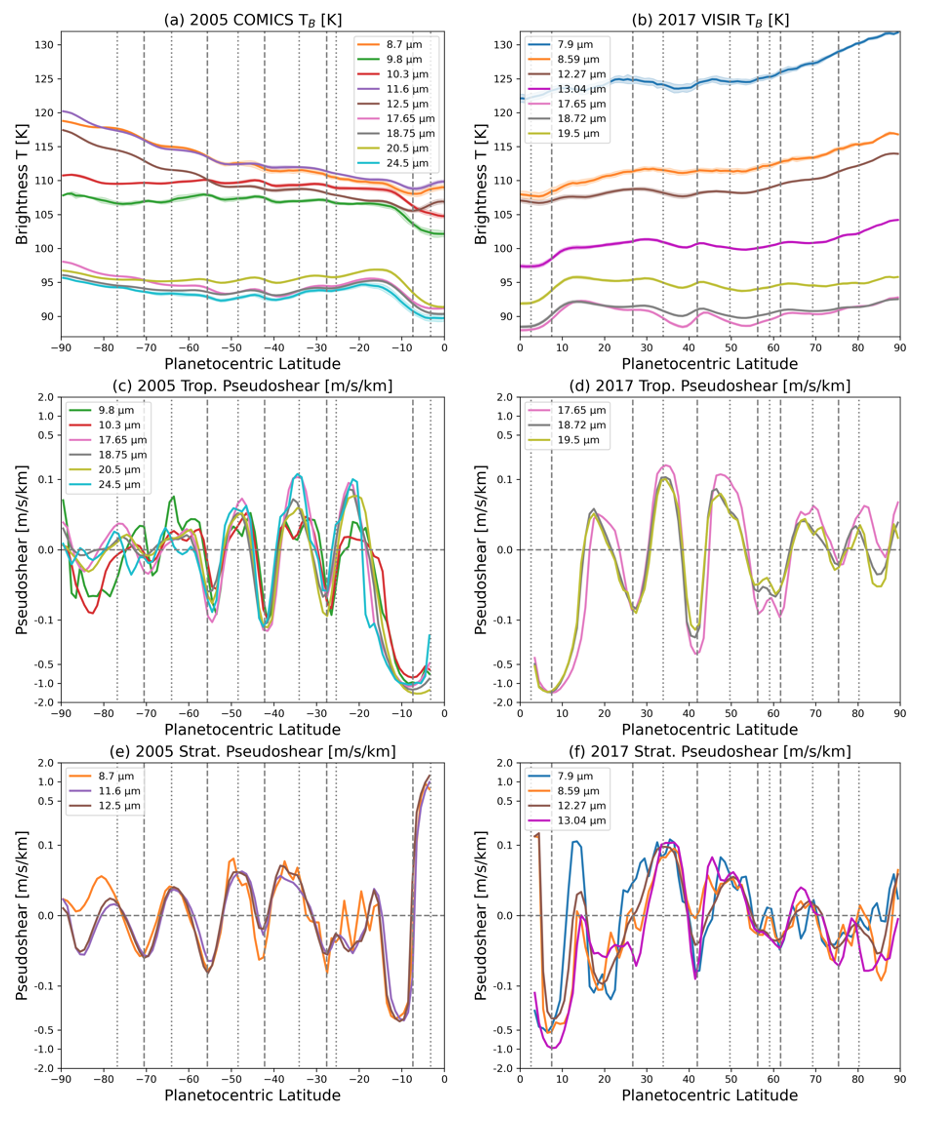}}
\caption{(a,b) Zonally-averaged brightness temperatures observed during southern summer (April 2005 by COMICS, left) and during northern summer (September 2017 by VISIR, right).  Vertical dashed lines indicate local maxima in the Cassini-derived wind field (eastward jets), vertical dotted lines indicate local minima (westward jets).  Shaded regions represent the standard deviation of the zonal average, but do not include radiometric calibration uncertainties. Centre-to-limb variation has been approximately removed as described in the text.  (c,d) Corrected pseudoshear for troposphere-sensing filters, estimated from the latitudinal $T_B$ gradients as described in the main text, for 2005 and 2017, showing the correspondence of Saturn's thermal gradients and the wind field. (e,f) show corrected pseudoshear for stratosphere-sensing filters. Note that the y-axis scale is linear within $\pm0.1$ m/s/km, and logarithmic thereafter.  Pseudoshears are not shown within $\pm3^\circ$ of the equator, where the Coriolis parameter $f$ tends to zero.}
\label{zonalshear}
\end{centering}
\end{figure*}

We used the highest-quality observations from both dates to generate calibrated equirectangular maps.  A crude correction for limb-darkening/brightening effects was derived by sampling the brightness at all emission angles ($\mu=\cos\theta>0.4$) within a restricted latitude range $\pm10^\circ$ around the sub-observer latitude.  This centre-to-limb effect was fitted with a fourth-order polynomial, which was then used as an empirical correction to the maps to remove the emission-angle dependence.  An estimate of the zonal mean was derived by averaging within $\pm30^\circ$ of the central meridian in each filter, and these are shown in Fig. \ref{zonalshear} for both epochs.  

To correct for the latitudinal dependence of the brightness temperature $T_B$ on local gravity $g$ and the Coriolis parameter $f=2\Omega\sin\theta$, we followed \citet{21fletcher} in estimating a `pseudo-shear` $\Delta$ on constant-pressure surfaces:
\begin{equation}
    \Delta=-\frac{g}{fT_B}\frac{\partial T_B}{\partial y} 
\end{equation}
where $y$ is the north-south distance.  If we assume that the brightness temperature provides a good approximation to the kinetic temperature of the atmosphere at the peak of the contribution function for each filter, then this $\Delta$ can be viewed as a crude estimate of the vertical shear on the zonal winds, $du/dz$.  The strong seasonal temperature contrasts in Saturn's troposphere and stratosphere produce significant trends in $\Delta$ as a function of latitude \citep[such seasonal changes in windshear are quite valid and have been discussed elsewhere, e.g.,][]{12friedson}.  However, the purpose of this crude calculation is to identify the edges of Saturn's thermal bands, rather than to provide a quantitative estimate of the zonal winds at all altitudes, so we removed these seasonal trends via linear fitting between $15-70^\circ$ in each hemisphere, and subtracting it from the measured $\Delta$.  This adjusted form of $\Delta$ is shown in Fig. \ref{zonalshear}c-f, and is observed to oscillate back and forth around zero, giving a good indication of the boundaries of the thermal bands observed in the VISIR and COMICS images.  We caution the reader that $\Delta$ should never be used quantitatively for zonal wind calculations as it lacks the seasonal temperature gradients, and that latitudes away from the sub-observer point will be viewed with lower spatial resolutions, such that $\Delta$ should be interpreted only as a qualitative guide to the edges of Saturn's thermal bands.

\subsubsection{Temperate Mid-Latitudes}

In Fig. \ref{zonalshear}c-f, we show how $\Delta$ compares to the peaks of the eastwards (prograde) and westwards (retrograde) cloud-top winds as measured by Cassini/ISS \citep{11garcia}.  Focusing first on the \textit{temperate mid-latitudes} ($30-70^\circ$ latitude in each hemisphere), Fig. \ref{zonalshear} suggests that westward jets are associated with $\Delta>0$ in all COMICS and VISIR filters, both in the troposphere (panel c-d) and stratosphere (panel e-f).  However, for the eastward jets (which tend to be narrower), there are examples of both positive and negative $\Delta$, although there is a clear trend for $\Delta<0$ to be associated with strong eastward jets at mid-latitudes.  Cassini/CIRS had previously shown this in the troposphere with true windshears of a similar magnitude \citep[Fig. 16d of ][]{16fletcher}, but the stratosphere-sensing ground-based measurements at 7.9, 8.6-8.7, 12.3-12.5, and 13.0 $\mu$m in Fig. \ref{zonalshear}e,f reveal that this close correspondence also exists in the stratospheric brightness temperatures, indicating that the banded structure extends from the 500-mbar level to at least 1-10 mbar.  

\begin{figure*}
\begin{centering}
\centerline{\includegraphics[angle=0,scale=.3]{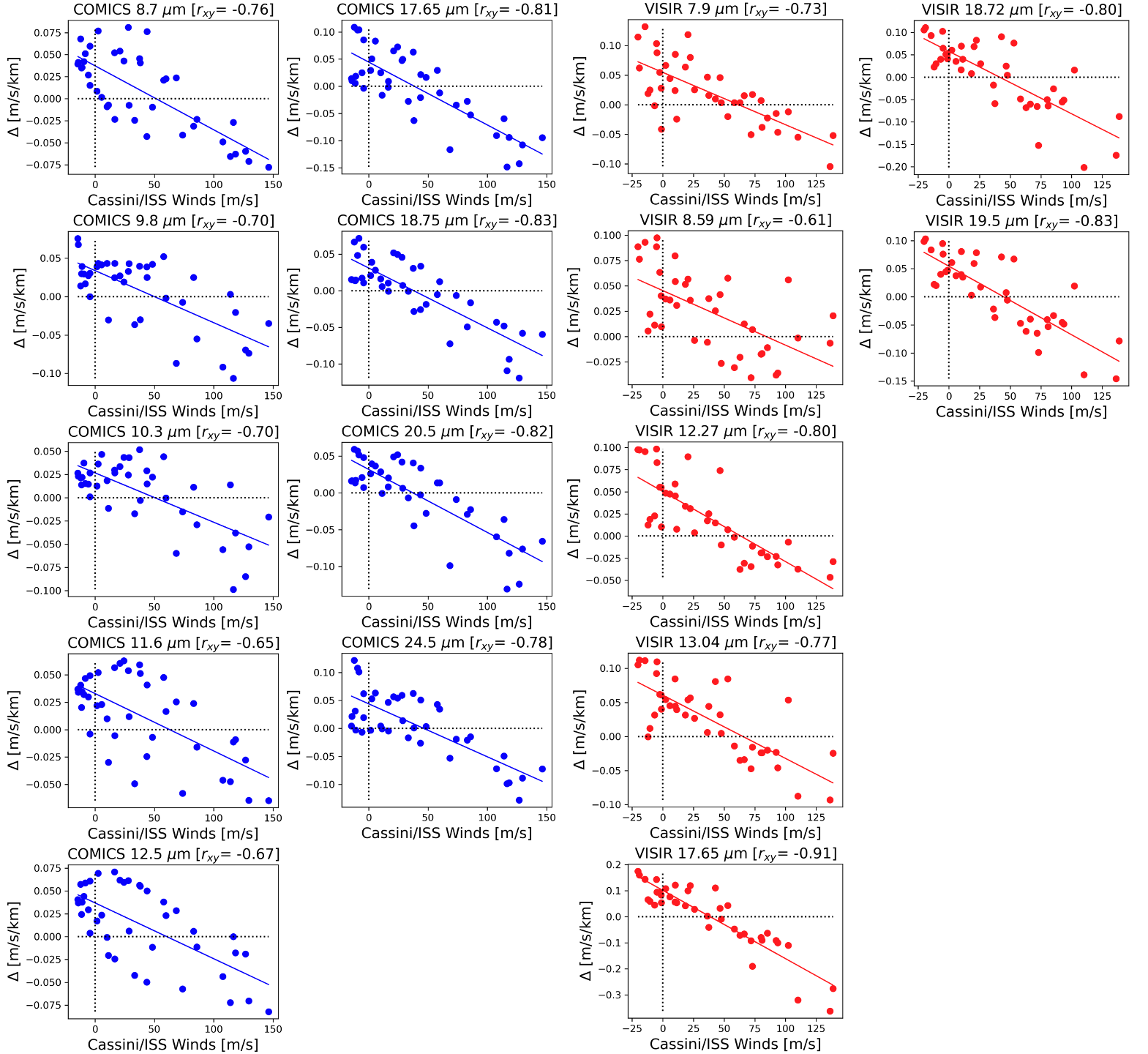}}
\caption{Negative correlation between Cassini-derived cloud-top winds of \citet{11garcia} and the pseudoshear $\Delta$, measured in the $30-70^\circ$ latitude range for each hemisphere.  Left columns show the correlation during southern summer in blue (April 2005, COMICS), right columns show the correlation during northern summer in red (September 2017, VISIR).  Pearson correlation coefficients $r_{xy}$ are provided for each panel, along with a linear regression line.}
\label{figure:9 correlation}
\end{centering}
\end{figure*}

We assessed the strength of the correlation by calculating the Pearson correlation coefficient $r_{xy}$ and their associated $p-$values, as shown in Fig. \ref{figure:9 correlation}. We focused on Saturn's temperate mid-latitudes between $30-70^\circ$ so as to avoid bias from the strong equatorial jet (see below), and poor sampling of the polar regions.  Moderate negative correlation between winds and the corrected pseudoshear of $r_{xy}=0.75\pm0.1$ was observed in all filters, confirming that all mid-latitude zonal jets experience pseudoshear opposing their cloud-top motions. Fig. \ref{figure:9 correlation} suggests that this correlation is generally stronger in the troposphere-sounding filters than the stratosphere-sounding 7.9- and 11.6-to-12.5-$\mu$m filters, but nevertheless remains significant in all cases.  If this pseudoshear, which is based on brightness temperature rather than kinetic temperature, is a good estimate of the true windshear at the altitudes probed by each filter, then this confirms the weakening of the winds with altitude continues from the troposphere into the stratosphere.  

The strong correlation between Saturn's mid-latitude jets and the latitudinal temperature gradients suggests that the banded structure of belts and zones is better defined by temperature and wind mapping than by the visible albedo \citep[e.g.,][]{06vasavada}.  As on Jupiter, Saturn's eastward jets separate cool anticyclonic zones on their equatorward sides from warm cyclonic belts on their poleward sides, and this pattern persists throughout Saturn's temperate latitudes in both hemispheres. Evidence from aerosols and gaseous tracers, particularly NH$_3$, PH$_3$ and para-H$_2$ \citep[reviewed by][]{20fletcher_beltzone} suggest upwelling in zones and subsidence in belts, such that each zonal jet may be associated with a meridional circulation cell akin to the Earth's Ferrel cells \citep{09delgenio, 20fletcher_beltzone}, a phenomenon that is also at work on Jupiter \citep[e.g.,][]{21duer}.  The VISIR and COMICS data presented here show how ground-based imaging can reveal thermal contrasts associated with Saturn's meridional circulation patterns on the scale of the zonal jets.

\subsubsection{Tropical Latitudes}

The correspondence between tropospheric and stratospheric $\Delta$ becomes less clear at \textit{tropical latitudes} within $\pm30^\circ$ of the equator, where Saturn's wind field is dominated by the broad eastward jet \citep{11garcia}.  For example, the eastward jet at $27\circ$N coincides with $\Delta<0$ for troposphere-sounding filters (13.0-19.5 $\mu$m in Fig. \ref{zonalshear}c,d), but $\Delta>0$ for stratosphere-sounding filters (7.9, 8.6 and 12.3 $\mu$m, Fig. \ref{zonalshear}e,f), indicating that $\Delta$ changes direction at altitudes above the tropopause.  Intriguingly, this is not seen for the mirror-jet at $29^\circ$S, where all the images have $\Delta<0$.  Another example are the small maxima in the winds at $\pm7.5^\circ$ (planetocentric), which are superimposed onto Saturn's broad equatorial jet.  These are associated with strong negative pseudoshear near $\pm10^\circ$ in the troposphere-sounding filters, but the stratosphere-sensing filters are notably different in Fig. \ref{zonalshear}e,f.

These low-latitude pseudoshears are likely to be temporally variable, particularly in the stratosphere.  In 2005 and 2017, $\Delta$ remained negative for the troposphere-sensing filters (17-19 $\mu$m) throughout the $0-15^\circ$ region (i.e., the equatorial zone is cold).  Conversely, the stratospheric filters (8.7, 11.6 and 12.5 $\mu$m for COMICS; 7.9, 8.6, 12.27 and 13.04 $\mu$m for VISIR) show positive $\Delta$ equatorward of $7.5^\circ$ in 2005 (Fig. \ref{zonalshear}e), but remain negative in 2017 (Fig. \ref{zonalshear}f).  This difference is likely associated with the equatorial contrasts driven by Saturn's stratospheric oscillation.  The equatorial maximum was stronger in 2005 than in 2017, but as only $\sim12$ years had elapsed between these observations, compared to the $\sim15$-year period of Saturn's equatorial oscillation \citep{08orton_qxo}, we may be at an earlier stage of the warming at 1-10 mbar.  This is consistent with the findings of \citet{17fletcher_QPO}, but see Section \ref{post_cassini} for discussion of this progression beyond 2017.

\subsubsection{Polar Latitudes}

Poleward of $70^\circ$ in both hemispheres, Fig. \ref{zonalshear}c-f shows weaker correspondence between $\Delta$ and the zonal jets.  Eastward jets at $71^\circ$S and $76^\circ$N (i.e., the latitude of Saturn's hexagon) show local minima in $\Delta$ in both the troposphere and stratosphere, and there are hints of $\Delta<0$ in the latitude domain surrounding the SPC in 2005 ($87^\circ$S) and the NPC in 2017 ($86^\circ$N), i.e., the strength of the cyclonic vortices decays with altitude.  Nevertheless, these polar gradients are more scattered than at mid-latitudes, likely because the magnitude of $\Delta$ is extremely sensitive to the spatial resolution of the polar $T_B$ measurements in Fig. \ref{zonalshear}a-b.

% It is most notable at low latitudes, where the small maxima in the winds at $\pm7.5^\circ$ (planetocentric), which are superimposed onto Saturn's broad equatorial prograde jet, are associated with strong negative pseudoshear near $\pm10^\circ$ in the Q-band filters sounding the upper troposphere.  

\subsection{Temperature Retrieval}
\label{inversion}

\begin{figure*}
\begin{centering}
\centerline{\includegraphics[angle=0,scale=.8]{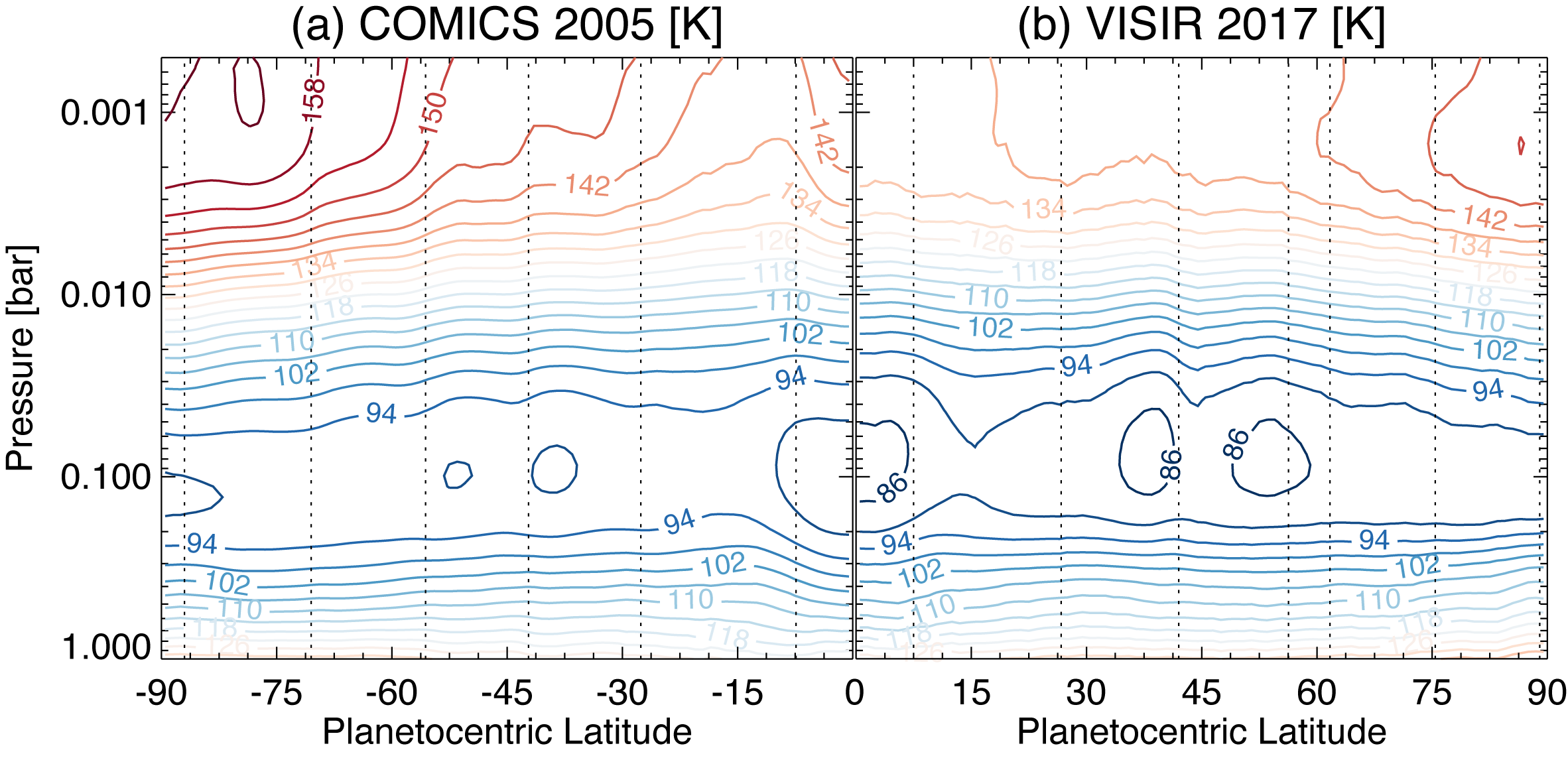}}
\caption{Temperature cross-sections during southern summer \citep[left, derived from COMICS in April 2005,][]{09fletcher_imaging} and northern summer (right derived from VISIR in September 2017).  Spectra formed from stacking the available filters in Fig. \ref{zonalshear} were inverted to provide a crude estimate of the temperature, with uncertainties ranging from $\sim1.5$ K at 500 mbar, to $\sim2.0$ K at 100 mbar, to $\sim3.5$ K at 1 mbar.  Vertical dotted lines show the locations of the prograde eastward jets at the cloud tops.}
\label{figure:10 Tretrieval}
\end{centering}
\end{figure*}

As a final demonstration of the capabilities of 8-m facilities to constrain the thermal structure of Saturn's atmosphere, we perform a crude spectral inversion of the April-2005 COMICS observations and September-2017 VISIR observations to estimate the 2D (latitude-pressure) temperature structure, in Fig. \ref{figure:10 Tretrieval}.  A single latitude-independent temperature profile \citep[a low-latitude average from][]{09fletcher_ph3} was used as the \textit{a priori} for retrievals at all latitudes and epochs. The COMICS inversion repeats that shown in Fig. 14 of \citet{09fletcher_imaging}, and includes the significant uncertainties inherent in retrievals from stacks of 7-9 narrow-band images spanning 7-25 $\mu$m.  This is an under-constrained problem, with a small number of spectral points being used to determine a smoothed $T(p)$ structure, using observations that have been calibrated by scaling to Cassini/CIRS.  Furthermore, the two inversions possess differing information content:  the 2005 COMICS inversion lacks a filter sensing stratospheric CH$_4$ at 7.9 $\mu$m \footnote{A 7.9-$\mu$m filter was later added to the COMICS instrument.}  The stratospheric temperatures were therefore primarily controlled by the COMICS 12.5-$\mu$m filter, which senses C$_2$H$_6$ \citep[a latitude-independent C$_2$H$_6$ vertical profile was used as the prior, based on][]{18fletcher_poles}.  Similarly, the VISIR inversion lacks deeper-sounding N-band observations at 9-11 $\mu$m, such that the 2017 tropospheric temperature are controlled only by the $13-19 \mu$m filters.

Formal retrieval errors range from $\sim1.5$ K at 500 mbar, to $\sim2.0$ K at 100 mbar, to $\sim3.5$ K at 1 mbar.  The absolute temperatures at each level are therefore rather uncertain, but Fig. \ref{figure:10 Tretrieval} does serve to show (i) the seasonal trends in hemispheric temperatures near southern and northern solstice; and (ii) the modulation of the temperatures by the belt/zone structures associated with the shear on the zonal winds (Section \ref{beltzone}).  At 1 mbar, the peak temperatures within the SPSV ($\sim162$ K) and the NPSV ($\sim150$ K) are within $\pm5$ K of the temperatures measured by Cassini/CIRS during the same periods \citep{18fletcher_poles}.  At 110 mbar, the temperature estimates during southern summer ($90\pm2$ K) and northern summer ($86\pm2$ K) are consistent with those derived from Cassini/CIRS \citep{18fletcher_poles}.  This is not surprising, given the use of CIRS data to calibrate the ground-based data, but demonstrates the self-consistency of this approach.

%%%%%%%%%%%%%%%%%%%%%%%%%%%%%%%%%%%%%%%%%%%%%%
%%%%%%%%%%%%%%%%%%%%%%%%%%%%%%%%%%%%%%%%%%%%%% 
%%%%%%%%%%%%%%%%%%%%%%%%%%%%%%%%%%%%%%%%%%%%%%

\section{Saturn's Seasons before Cassini}
\label{extension}

% Introduction to the longer-term time series.
The high-resolution ground-based record of Saturn's tropospheric and stratospheric temperatures described in Section \ref{cirs_comparison} spans slightly less than half a Saturnian year, from 2004 to 2017.   However, lower-resolution images of Saturn's mid-infrared emission \citep{89gezari, 08orton_qxo} extend back to the 1980s.  In particular, observations from NASA's Infrared Telescope Facility (IRTF) span the previous northern spring (June 1984, $L_s=50^\circ$, to December 1987, $L_s=90^\circ$), northern summer ($L_s=90-180^\circ$, until November 1995), and northern autumn ($L_s=180-270^\circ$, until October 2002).  In Section \ref{interannual} we present this longer-term time series, and generate synthetic images based on Cassini-era temperature and hydrocarbon distributions to search for evidence of interannual variability.  Then in Section \ref{radiative} we compare the mid-IR observations to numerical radiative-convective models \citep{14guerlet} and radiative-photochemical models \citep{16hue}.

\subsection{Inter-annual Variability}
\label{interannual}

\begin{figure*}
\begin{centering}
\centerline{\includegraphics[angle=0,scale=.8]{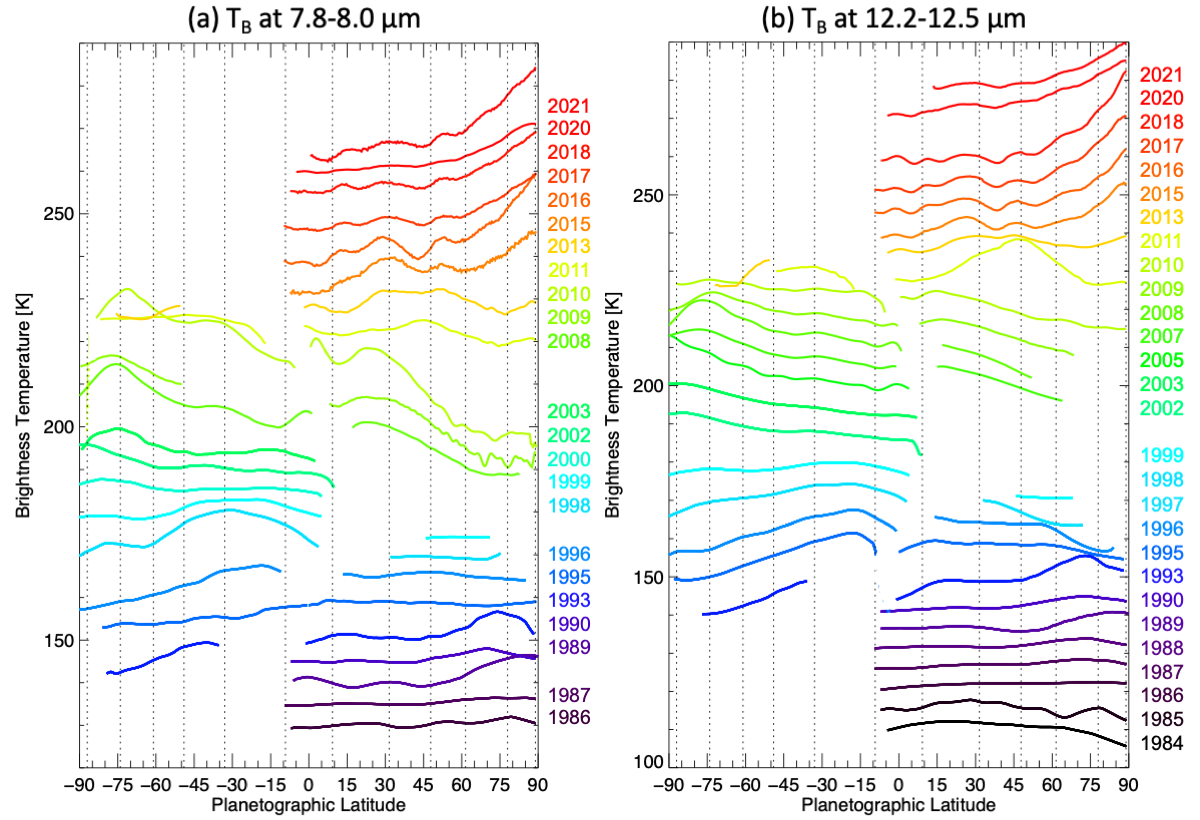}}
\caption{Zonal-mean stratospheric brightness temperatures (CH$_4$ emission near 8 $\mu$m, and C$_2$H$_6$ emission near 12.3 $\mu$m) extracted from selected observations from 1984 to 2021. Temperature profiles are offset from the earliest date by multiples of 5 K for clarity, and colour-coded by year. Dotted lines show the location of Saturn's eastward jets.  Standard deviations on zonal averages are less than 0.5 K, but this does not include systematic calibration uncertainties.  Regions of ring obscuration are omitted.  Given the significant improvement in image quality from 1984 to 2021, this comparison should be treated with caution, as described in the main text.  In particular, the bright polar emissions in the 1980s datasets often appear near $\sim75^\circ$N rather than right at the pole, due to the coarse spatial resolution of the earliest datasets.}
\label{waterfall}
\end{centering}
\end{figure*}

Fig. \ref{waterfall} displays annual zonal-mean brightness temperatures for two filters sensing stratospheric CH$_4$ (7.8-8.0 $\mu$m) and C$_2$H$_6$ (12.2-12.5 $\mu$m). This figure extends that of \citet{08orton_qxo}, and the averages are extracted from a longitude range within $\pm30^\circ$ of the central meridian from images such as those shown in Figs. \ref{figure:1 thermal_images_comics_visir_pt1}-\ref{figure:4 thermal_images_MIRLIN_MIRSI}.  As the quality of the mid-IR imaging has improved significantly over the past four decades, a quantitative comparison of the observations is challenging:  for example, the visibility of the banded structure and warm polar vortices varied considerably as spatial resolution and detector sensitivity has improved.  Lower spatial resolutions blend Saturn's limb emission with the surrounding dark sky, artificially darkening the limb and moving the brightest polar emission to lower latitudes (e.g., the peak often occurs near $75^\circ$N in the 1984-1993 datasets in Fig. \ref{waterfall}).  Nevertheless, Fig. \ref{waterfall} displays qualitative trends:  (i) the seasonal progression of the hemispheric asymmetry, warming in the spring and cooling in the autumn; (ii) the peak emission from the summertime poles as the North/South polar stratospheric vortices (NPSV, SPSV) form \citep{18fletcher_poles}; and (iii) the change in low-latitude temperature contrasts associated with Saturn's equatorial stratospheric oscillation \citep{08orton_qxo}.  We shall return to each of these points below.

% /data/nemesis/jsdb3/visir/TB_7.9/plot_lat_tb.pro

\begin{figure}
\begin{centering}
\centerline{\includegraphics[angle=0,scale=.3]{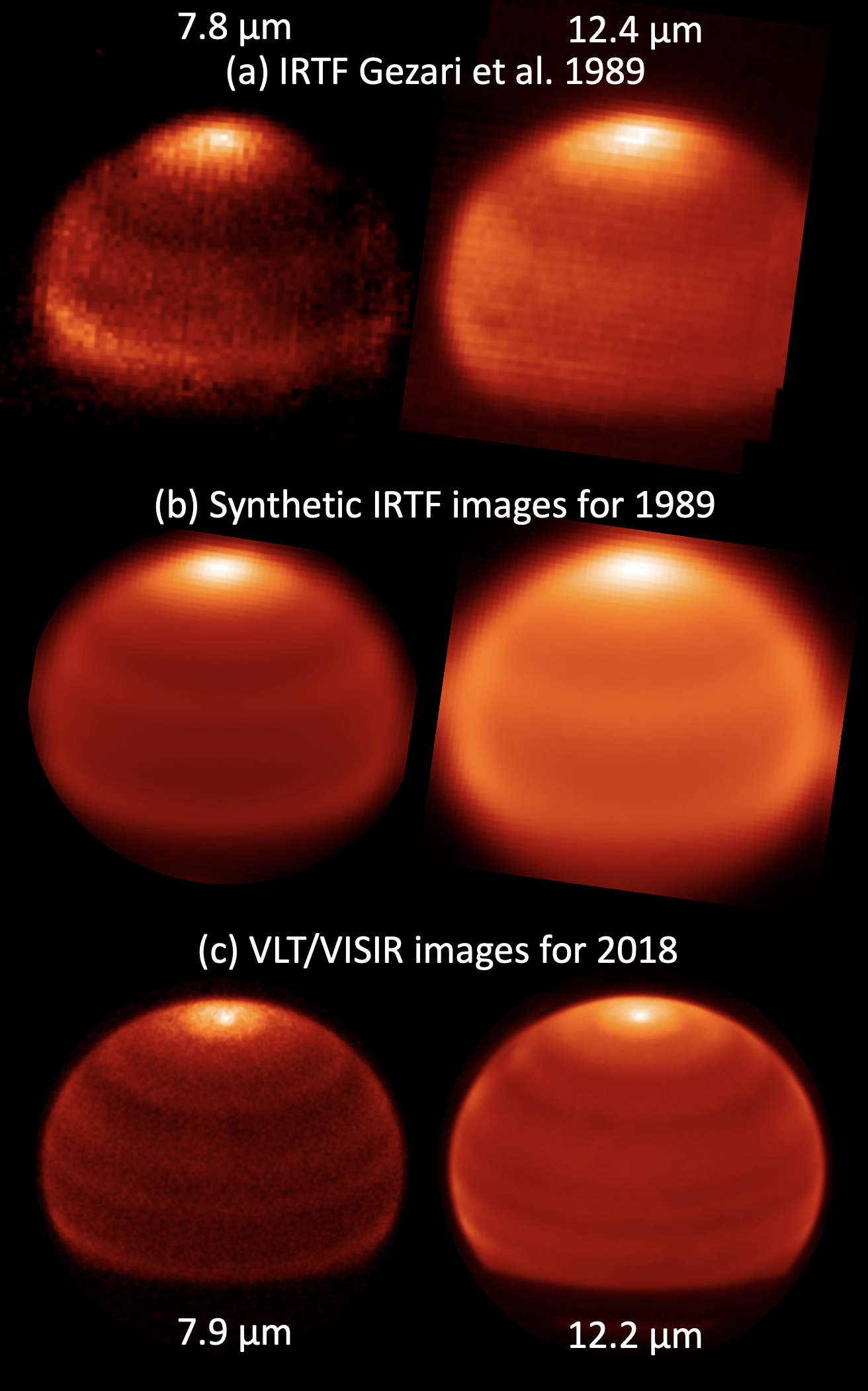}}
\caption{Comparison of synthetic images (b) based on the March-1989 ($L_s=104^\circ$) observations of \citet{89gezari} using a 3-m mirror (a).  These can be compared to VLT/VISIR observations from May 2018 ($L_s=101^\circ$), at approximately the same moment in Saturn's seasonal cycle.}
\label{gezari_synths}
\end{centering}
\end{figure}

Given the proposed semi-annual nature of the equatorial oscillation \citep{08orton_qxo, 08fouchet}, our first-order assumption would be that Saturn should appear the same from one Saturnian year to the next.  Thus temperatures measured by Cassini/CIRS near the end of the mission ($L_s=93^\circ$, September 2017) should provide a good approximation to the first 2D thermal images of Saturn observed one Saturnian year earlier \citep[March 1989, $L_s=103^\circ$,][]{89gezari}.  The exact anniversary of $L_s=103^\circ$ was August 2018, after the end of the Cassini mission.  Using the technique described in Section \ref{synth_images}, we generated synthetic images in Fig. \ref{gezari_synths} to match the observations of \citet{89gezari}, using the platescale and field-of-view of the 2D camera in 1989, but using the temperatures and hydrocarbon distributions from Cassini measurements in 2017.  The filter functions of the original IRTF camera are not available, so we used Gemini/Michelle filter functions as a substitute for the three filters \citep{97glasse}.   The synthetic images and data are shown in Fig. \ref{gezari_synths}, and are compared to VISIR observations in May 2018, close to the anniversary of the 1989 observations.

\begin{figure}
\begin{centering}
\centerline{\includegraphics[angle=0,scale=.4]{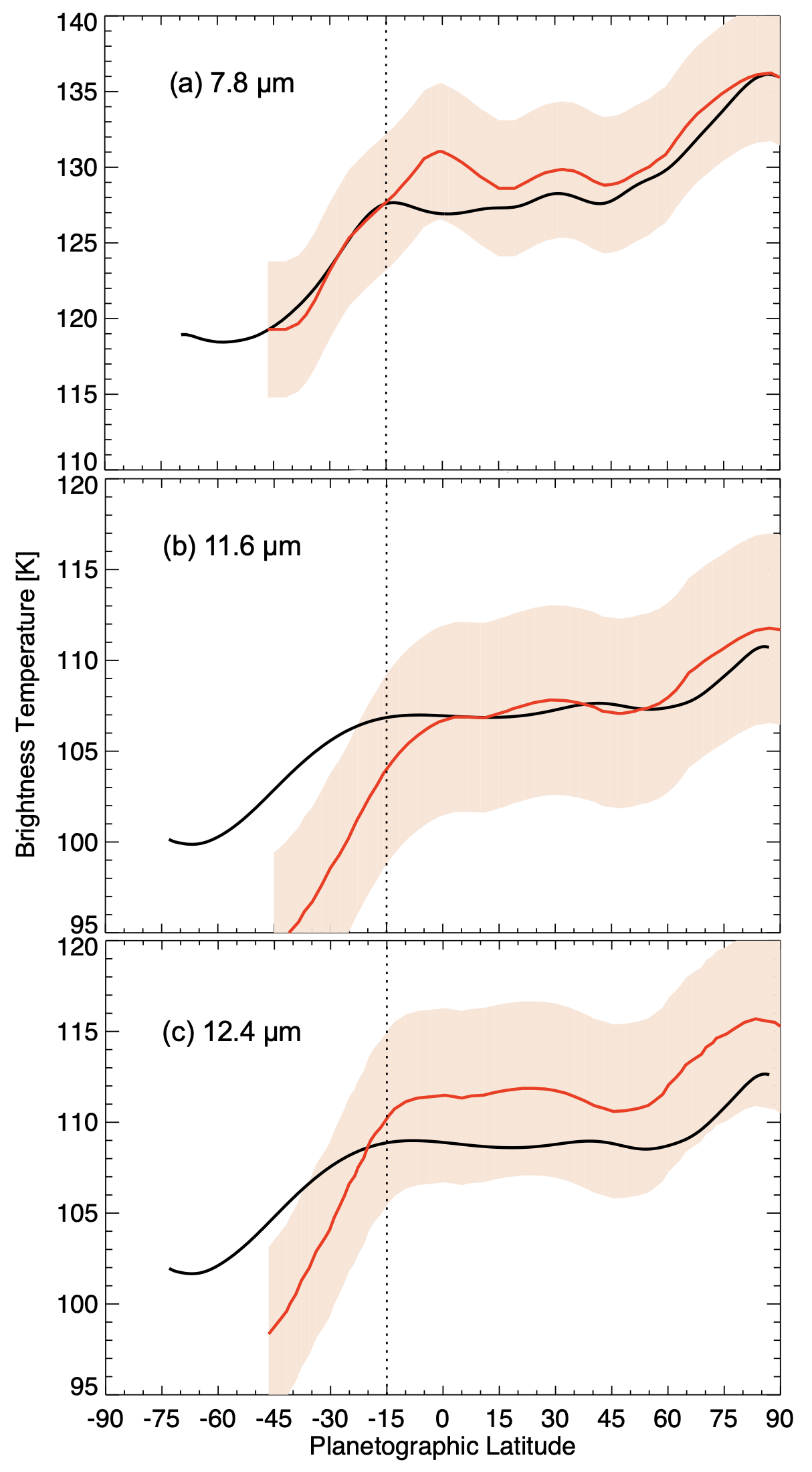}}
\caption{Comparison of Saturn's zonally-averaged brightness in March 1989 \citep[red curve,][]{89gezari} with scans extracted from synthetic images (black curve) at 7.8 $\mu$m, 11.6 $\mu$m and 12.4 $\mu$m. Regions southward of the vertical dotted line (black) are obscured by the rings.  Absolute uncertainties on the brightness temperatures (shaded regions) are estimates from the standard deviation of the background in the 1989 images. Although these are large compared to the measured zonal contrasts, the relative variability is consistent with that shown in Fig. \ref{gezari_synths}(a).}
\label{figure:12 interannual_variability}
\end{centering}
\end{figure}

Fig. \ref{figure:12 interannual_variability} directly compares the synthetic images with the calibrated 1989 observations at 7.8 $\mu$m, 11.6 $\mu$m and 12.4 $\mu$m.  We sampled the brightness temperatures through the same longitude as \cite{89gezari}; $20^\circ$ east of the central meridian. While the comparison is reasonable, with the warm polar domain $>70^\circ$N observed in all three filters, the detection of the same warm mid-latitude band near $30^\circ$N at 7.8 $\mu$m, and predicted temperatures within the range of uncertainty in the calibration of the \citet{89gezari}, there remain some structural differences.  The rapid drop in brightness temperature south of $15^\circ$S is due to absorption from Saturn's rings at 11.6 and 12.4 $\mu$m.  Ethane emission at 12.4 $\mu$m appears to be systematically brighter than the Cassini prediction in the northern hemisphere by $\sim4$ K, but this is likely due to challenges in filter calibration and uncertainties associated with the filter function, as temperature offsets are not seen at 7.8 and 11.6 $\mu$m.  However, close inspection of maxima and minima in the 11.6- and 12.4-$\mu$m latitude profiles suggest latitudinal offsets in the location of warm bands.  Mid-latitude differences between 2017 and 1989 could be real, given thermal and chemical consequences of the 2010 northern springtime storm \citep{12fletcher, 15moses}.  However, mid-latitude stratospheric perturbations from the storm had largely dissipated by 2017 \citep{17fletcher_QPO}, and the temperature-dependent photochemistry within the `beacon' should not have significantly affected the C$_2$H$_6$ abundance \citep{15moses}.  The apparent differences in the $15-60^\circ$N domain may not be genuine, given the difficulties of navigating the precise location of the limb in the 1989 images.  

The most robust difference between the 1989 observations and the Cassini-epoch predictions is at low latitudes at 7.8 $\mu$m.  Both the 1989 images in Fig. \ref{gezari_synths} and the zonal average in Fig. \ref{figure:12 interannual_variability} display a warm equatorial band at 7.8 $\mu$m.  This warm band is not present in our synthetic images in Fig. \ref{gezari_synths}b (based on 2017 CIRS data), nor in the May-2018 VISIR observations in Fig. \ref{gezari_synths}c. This warm equatorial band could also have been present in 7.8-$\mu$m images before 1989, but the low spatial resolution and high uncertainty on these early images in Fig. \ref{figure:3 thermal_images_GRABER_GEZARI} make it hard to be conclusive.  The warm band can be seen in the 1990 images in Fig. \ref{figure:3 thermal_images_GRABER_GEZARI}, albeit at low contrast because of the warmer belt immediately north of the equator.  This is consistent with Fig. 3 of \citet{08orton_qxo}, which suggests a peak in the equatorial brightness near northern summer solstice. 

Fig. \ref{waterfall}a reveals that the warm band was absent through the 1990s, but could be seen again in IRTF and Keck imaging in 2002-2004 \citep[Fig. \ref{figure:1 thermal_images_comics_visir_pt1},][]{05orton}, throughout 2007-2013 images from COMICS and VISIR in Fig. \ref{figure:1 thermal_images_comics_visir_pt1}, but COMICS and VISIR images between 2015 and 2020 (Fig. \ref{figure:2 thermal_images_comics_visir_pt1}) shows that the equator was cooler throughout this epoch.  Unlike the CH$_4$ sensing filter, a warm equatorial band has been observed at 12.2-12.5 $\mu$m throughout the 2007-2020 period, suggesting that this filter sounds a deeper level of the equatorial oscillation.

The local minimum in equatorial 7.8-$\mu$m brightness temperature near northern summer solstice in 2017 is also seen in the Cassini/CIRS brightness temperatures (Fig. \ref{CIRS-VISIR-Lat-tb-comparison}A), so it is no surprise that our Cassini-epoch predictions of 1989 brightness temperatures expects a local equatorial minimum in Fig. \ref{figure:12 interannual_variability}a.  If the equatorial stratospheric oscillation were truly semi-annual at this altitude (10 mbar), then we would expect the oscillation to display the same latitudinal contrasts one Saturnian year apart.  Saturn's tropical stratosphere therefore displays inter-annual variability, as the equatorial stratospheric oscillation is not semi-annual.  The $\sim15$ year period has been questioned previously:  \citet{14sinclair} found that 2.1-mbar temperatures derived from Cassini/CIRS in 2009-10 showed a local temperature maximum (consistent with our VISIR images in Fig. \ref{figure:1 thermal_images_comics_visir_pt1}), whereas those derived from Voyager/IRIS in 1980 (one Saturnian year earlier) showed a local temperature minimum.  The sinusoidal oscillation envisaged in Fig. 3 of \citet{08orton_qxo} does not seem to have held after $\sim2006$, meaning that equatorial temperature contrasts from the Cassini epoch are not a good model for the brightness in the 1980s.  

% New paragraph
Natural variability in equatorial oscillations is not unexpected.  No terrestrial equatorial oscillation, neither the quasi-biennial oscillation (QBO) nor the semi-annual oscillation (SAO), has a perfectly fixed period, and Earth's QBO has been disrupted in the recent past \citep{16osprey}.  Saturn's equatorial oscillation was disrupted for a period of $\sim3$ years, when a large warm stratospheric vortex was present, related to the northern springtime storm \citep{17fletcher_QPO}.  The contemporaneous nature of these two phenomena suggested momentum transfer by waves from Saturn's mid-latitudes to the equator, but this has yet to be confirmed via numerical modelling.  Jupiter's equatorial oscillation also appeared to shift periods as a consequence of global-scale changes in the planetary bands \citep{20giles_qqo, 21antunano}.  Thus equatorial oscillations can vary with time due to variable activity in the troposphere, such as extreme weather events or other wave-generating phenomena, which alter the wave-mean flow interactions.  The period of these oscillations also vary with height, as the vertical propagation of the waves that force the equatorial winds is not always the same, meaning that the altitude of wave-breaking constantly changes.  Saturn's equatorial oscillation could therefore be viewed as a hybrid of a QBO-like phenomenon (i.e., the descent of zonal shear zones towards the tropopause, driven by variable activity in the troposphere) and a seasonally-forced phenomenon, where the phase of the oscillation is naturally variable.  Furthermore, the time span of our observations remains insufficient to properly define the `mean' period and properties of Saturn's oscillation.

Considering Fig. \ref{waterfall}a and the images in Figs. \ref{figure:1 thermal_images_comics_visir_pt1}-\ref{figure:4 thermal_images_MIRLIN_MIRSI}, which show warm 7.9-$\mu$m bands evident in 1989-1990 and 2004-2010, and minima evident in 1996-2000 and 2015-2020, might suggest that a longer-term oscillation with a $\sim20$-year period could be more appropriate than the $\sim14-16$ years typically quoted.  This would help to explain why observations taken one Saturnian year apart are not in phase with one another, as shown by this work and \citet{14sinclair}.  However, VISIR observations in 2021 and 2022 indicate the re-emergence of the warm 7.9-$\mu$m band, $\sim15-17$ years after its last peak in 2004-2006, so the period remains considerably uncertain.  The existence of the local equatorial minimum in Fig. \ref{figure:2 thermal_images_comics_visir_pt1} between 2015 and 2020 could be a continued consequence of Saturn's 2010 storm, but this cannot explain why IRIS (1980) and CIRS (2009) observations near equinox were apparently out of phase.  The September 1990 storm \citep{91sanchez} is unlikely to be the cause, given that the equatorial oscillation appeared to be well-behaved between 1990-2005 in the study of \citep{08orton_qxo}.  Continued long-term monitoring of Saturn's equatorial oscillation will be needed to better constrain its periodicity, and whether the warm equatorial 7.9-$\mu$m band continues to re-establish itself in the years to come.

\subsection{Comparison with Radiative Models}
\label{radiative}

With the exception of the variability associated with the equatorial oscillation, and the strong emission from the northern storm `beacon' in 2011-2014, the $\sim37$-year time series in Fig. \ref{waterfall} suggests a slow seasonal reversal in hemispheric contrasts associated with radiative heating and cooling.  In this section we compare the observations with predictions based on the radiative climate models of \citet{14guerlet} and \citet{16hue}.

% Explanation of SPICE calculations and forward model
SPICE software tools provided by the Navigation and Ancilliary Information Facility \citep[NAIF,][]{96acton_naif}, accessed via `SpiceyPy' \citep{20annex}, were used to calculate Saturn's ephemerides at all oppositions (as viewed from Earth) between 1980 and 2030.  For each opposition, visible latitudes, longitudes, illumination and viewing angles were projected onto the sky using the 0.045 arcsecond-per-pixel grid of VISIR.  These were extracted along the central meridian on a 0.25" grid. For each latitude and $L_s$, we interpolated and extracted the relevant $T(p)$ from the model of \citet{14guerlet}, or the relevant $T(p)$ and ethane/acetylene distributions from the model of \citet{16hue}.  We compare the results from both models, as they used slightly different assumptions:  \citet{14guerlet} hold hydrocarbon distributions fixed with time, whereas \citet{16hue} iterate between a radiative model and photochemical model, allowing the distribution of the stratospheric coolants to change with time.  These subtle differences in hydrocarbon distributions have a secondary impact on the predicted brightness temperatures, which are primarily governed by the modelled stratospheric temperatures. 

\begin{figure*}
\begin{centering}
\centerline{\includegraphics[angle=0,scale=.9]{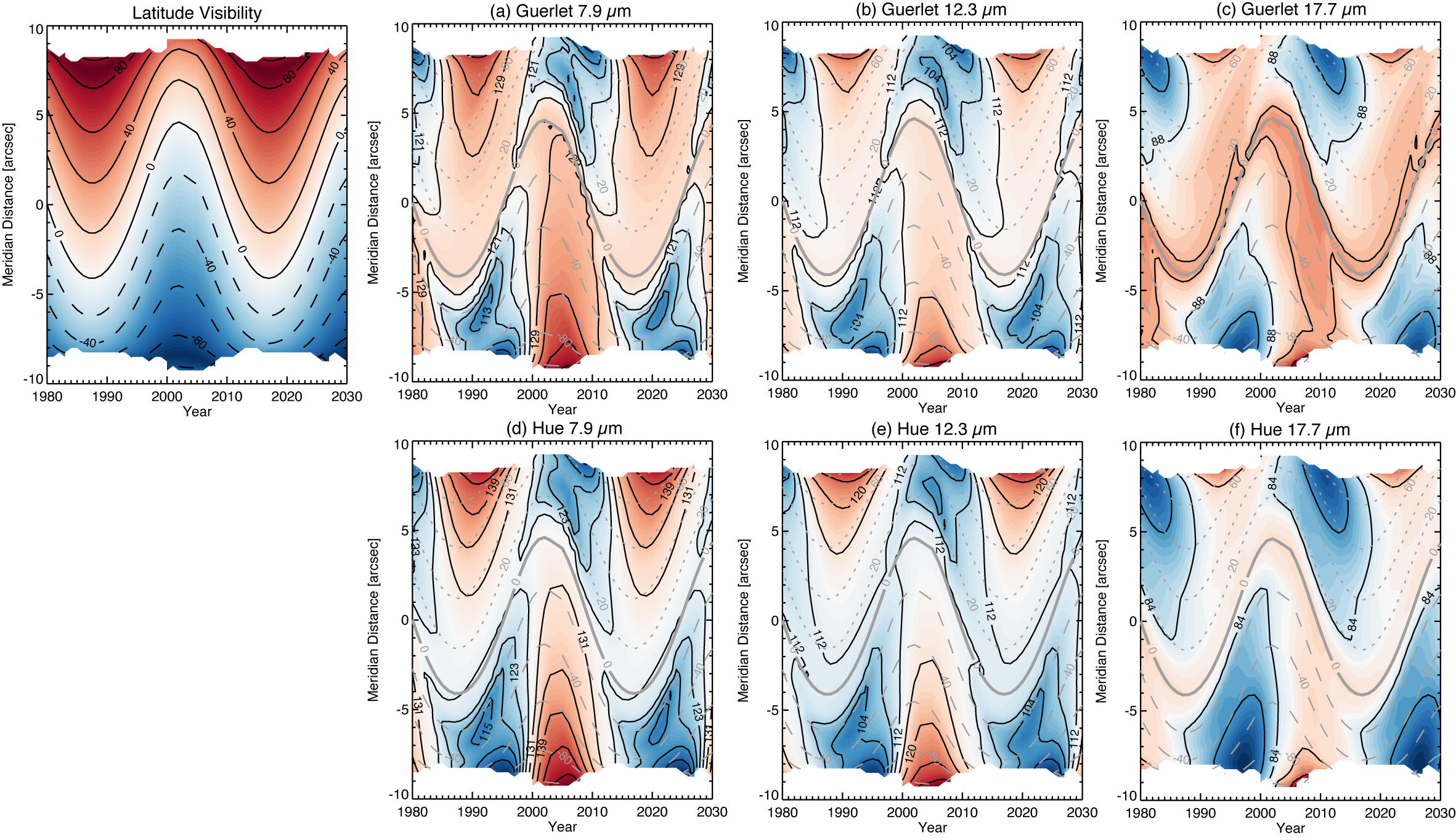}}
\caption{Model-based synthetic brightness temperatures (K) along Saturn's central meridian, calculated for opposition in each year from 1980 to 2030.  Values for $T_B$ are given by the solid black contours in panels (a)-(f) Three representative filters are shown (7.9, 12.3 and 17.6 $\mu$m), for the radiative models of \citet{14guerlet} (top row) and \citet{16hue} (bottom row).  Brightnesses are estimated on a 0.25" grid, with the zero-point being the centre of Saturn's disc.  Visible latitudes are shown in the top left (colours represent latitudes from blue in the south to red in the north), and the same latitudes are superimposed in grey on the brightness temperatures in $20^\circ$ steps (solid grey for the equator, dotted grey for northern latitudes, dashed grey for southern latitudes).  }
\label{figure:11 radmodel}
\end{centering}
\end{figure*}

% Discussion of Model Issues
Spectral radiance was forward-modelled in each VISIR filter (7.9-20 $\mu$m) for each central meridian position and date, providing a prediction of the brightness temperature variation as a function of time.  The results are shown in Fig. \ref{figure:11 radmodel} for three filters:  7.9 $\mu$m sensing stratospheric CH$_4$, 12.3 $\mu$m sensing stratospheric C$_2$H$_6$, and 17.7 $\mu$m sensing tropospheric H$_2$. Ring-obscured latitudes have not been removed. It is important to note that neither model was designed to produce the fine-scale belt/zone structure visible in Saturn's atmosphere, nor the pattern of equatorial stratospheric variability \citep{08orton_qxo, 17fletcher_QPO}.  Systematic discrepancies in the absolute temperatures predicted by the models and measured by CIRS are discussed extensively by the authors \citep{14guerlet, 16hue}.  Nevertheless, the radiative models are remarkably consistent with one another, and provide a useful prediction of the timings of peak temperatures (and hydrocarbon abundances).  The forward models in Fig. \ref{figure:11 radmodel} simply translate these into predicted brightnesses, based on the visibility and emission angles for each latitude.

%  These can \hl{compared to brightness temperatures extracted from the measurements in Fig.} \ref{figure:3 thermal_images_GRABER_GEZARI} and \ref{figure:4 thermal_images_MIRLIN_MIRSI}.  

% Comparison of stratospheric filters.
Fig. \ref{figure:11 radmodel} reveals that both models reproduce the observed brightening of the summertime poles, and that this peak brightness is offset from the solstices.  At 7.9 $\mu$m, the temperatures are expected to be relatively uniform along the central meridian in $\sim1983$ and $\sim2013$ in the northern hemisphere, and $\sim1998$ in the southern hemisphere.  A gradient in brightness then becomes established (the pole being warmer than the sub-observer point), peaking around 1991 and 2020 for the north pole in both models.  This peak brightness is 3-4 years after summer solstice, due to (i) the lagged warming of the stratosphere predicted by the models, and (ii) the fact that the warmest polar latitudes come close to the limb after solstice, their brightness being enhanced by limb brightening.  Furthermore, both models predict a warmer southern-summer peak 7.9 $\mu$m emission than for the northern summer, as a consequence of closer proximity to the Sun (perihelion).  The difference in angular size of Saturn from aphelion to perihelion can be readily seen in Fig. \ref{figure:11 radmodel}.  With ground-based observations it is then hard to track the subsequent cooling, as the high autumnal latitudes appear to recede from view faster than the predicted polar cooling.

% Comparison of older data
Both models predict that the north polar region should be brightest in 2020-21 in all three filters, and that future measurements should show a decline.  But what about the previous Saturnian summers?  The brightness contrast for the north polar region can be seen in Fig. \ref{waterfall}.  \citet{89gezari} should have been taking observations during the warmest epoch for the north pole in 1989-1990, as shown in Fig. \ref{figure:3 thermal_images_GRABER_GEZARI}, and the warm NPSV remained visible in 1993 and 1995 MIRAC2 observations in Fig. \ref{figure:4 thermal_images_MIRLIN_MIRSI}.  Observations by MIRLIN and SPECTROCAM10 in 1996 ($L_s=190^\circ$, shortly after equinox) reveals the absence of any bright polar emission at 7.9 and 12.3 $\mu$m.  This is counter to the expectations of both models in Fig. \ref{figure:11 radmodel}, which suggest that a weak positive gradient from the disc-centre to the north pole should persist to the late 1990s.  The data imply that the north polar region cools faster than predicted by either model - we speculate that enhanced radiative cooling by aerosols, photochemically produced during the maximum illumination of summer, might account for this discrepancy.  However, this could also be a consequence of poorer spatial resolution during the 1990 observations of northern autumn, as the VISIR observations of southern autumn (Fig. \ref{figure:1 thermal_images_comics_visir_pt1}) show the persistence of bright south-polar emission until 2010, consistent with the radiative models.

% Comparison of springtime conditions
What about the onset of polar brightening in Saturnian spring?  Both the 12.4-$\mu$m data in Fig. \ref{figure:3 thermal_images_GRABER_GEZARI} and the model suggest the first occurrence of bright north polar emission between 1984-1985 ($L_s=45-70^\circ$).  During the Cassini epoch, the north polar emission is predicted to have become visible after $\sim2014$ ($L_s\sim60^\circ$), within the 2011-15 gap in our time series between Fig. \ref{figure:1 thermal_images_comics_visir_pt1} and \ref{figure:2 thermal_images_comics_visir_pt1}, but consistent with the findings from Cassini/CIRS \citep{18fletcher_poles}.  During southern spring, the first hints of a warm south pole were observed at 7.8 $\mu$m by MIRLIN in 2000 ($L_s=233^\circ$, Fig. \ref{figure:4 thermal_images_MIRLIN_MIRSI}), consistent with the predictions of the radiative model.  This bright south polar emission was also observed by Keck in 2004, just prior to Cassini's arrival \citep{05orton}.   

% Can we tell the difference between the two models?
Given that \citet{16hue} also consider photochemical variations with time, we might have expected larger differences between the two models in Fig. \ref{figure:11 radmodel}.  Stronger latitudinal gradients in 7.9- and 12.3-$\mu$m emission are present in the \citet{16hue} model, and although these are closer to the observations, the absence of circulation makes the comparison challenging.  Furthermore, there is a subtle shift in the timing of the maximum brightness by approximately 1 year, closer to solstice for the model of \citet{16hue}.  Neither the 1985-95 nor the 2015-2020 datasets have sufficient temporal sampling to reliably distinguish between these models.  We return to the question of variability in the 2015-2022 observations in Section \ref{post_cassini}.

% Success in the troposphere
Finally, we note that neither model does a good job at reproducing the tropospheric brightness.  Both predict an equatorial maximum temperature following the annually-averaged insolation, whereas in fact the observed temperatures in Figs. \ref{figure:1 thermal_images_comics_visir_pt1} and \ref{figure:2 thermal_images_comics_visir_pt1} are more uniform, albeit with the belt/zone variability superimposed on top.  Nevertheless, the radiative models predict the existence of brighter polar emission at 17.7 $\mu$m that is consistent with the data, which show a brighter south polar emission between 2003-2009, and north pole between 2017-2020, and in 1990.  The data from 8-m facilities in Fig. \ref{figure:1 thermal_images_comics_visir_pt1}-\ref{figure:2 thermal_images_comics_visir_pt1} show this brightness to be entrained poleward of approximately $\pm75^\circ$ latitude, along with contributions from the compact polar cyclones.  

In summary, despite the problems at low latitudes and systematic differences in temperature, the models consistently reproduce the bright polar emission observed in the data, correctly predicting the onset of bright emission in springtime, and the disappearance of the bright emission in autumn.

%However, tentative evidence from the 1990s suggest that the NPSV cooled faster than radiative predictions as northern autumn approached, something we hope to test in the 2020s.

%%%%%%%%%%%%%%%%%%%%%%%%%%%%%%%%%%%%%%%%%%%%%%
%%%%%%%%%%%%%%%%%%%%%%%%%%%%%%%%%%%%%%%%%%%%%% 
%%%%%%%%%%%%%%%%%%%%%%%%%%%%%%%%%%%%%%%%%%%%%%

\section{Saturn after the Cassini mission}
\label{post_cassini}

\begin{figure}
\begin{centering}
\centerline{\includegraphics[angle=0,scale=.7]{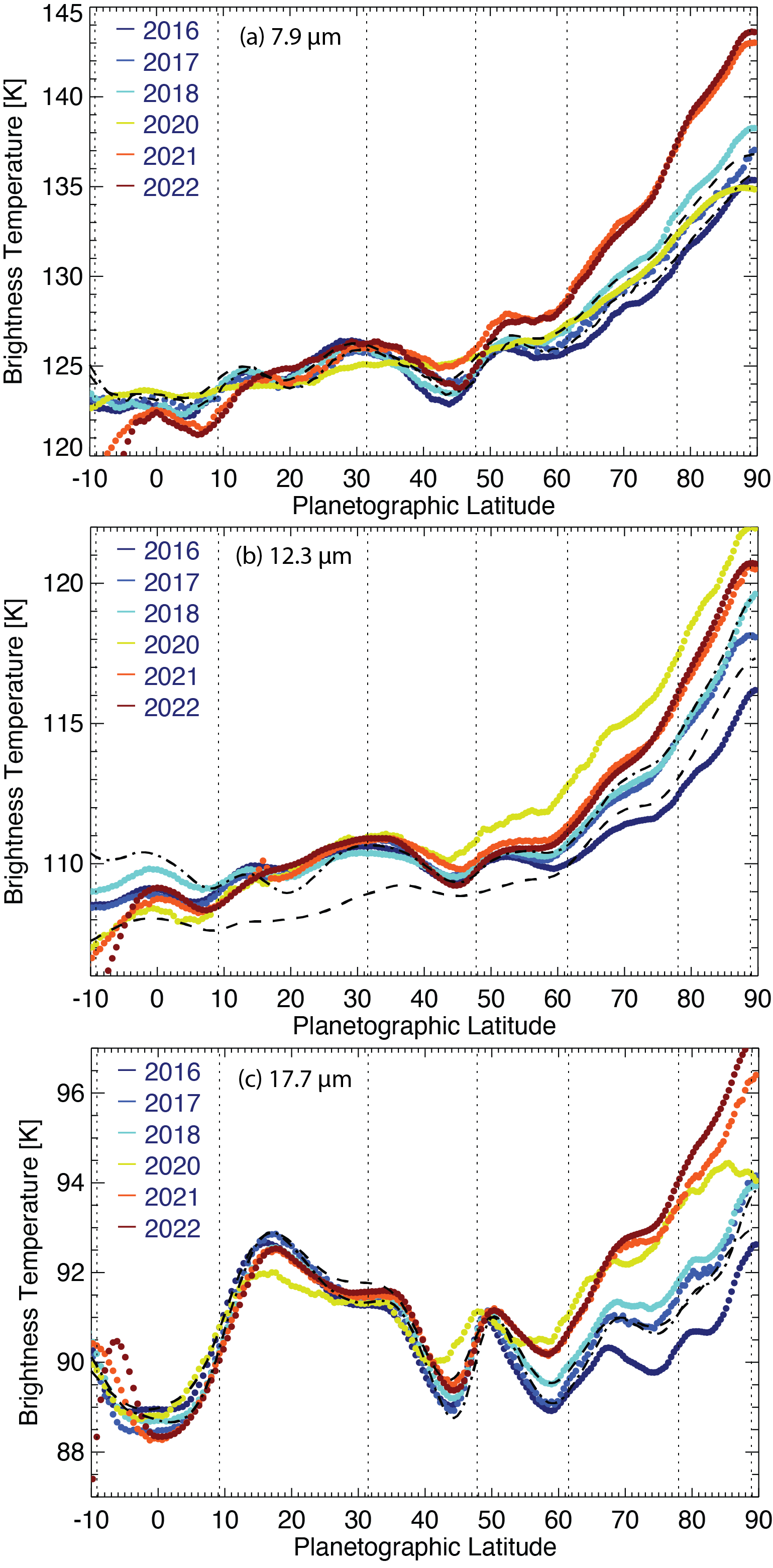}}
\caption{Saturn brightness temperatures in three filters for the years following the end of the Cassini mission.  Data are shown from VISIR in 2016, 2017, 2018, 2021, and 2022, showing a clear trend of warming poleward of $45^\circ$N.  Observations from COMICS in July 2020 are added, to plug a gap in the time series, but we note that COMICS used slightly different filters and a coarser platescale, making it hard to compare quantitatively to VISIR.  All have been calibrated to match CIRS brightnesses between $15-45^\circ$N in 2017.  Synthetic brightness temperatures, based on those measured in 2017 by Cassini, are shown for 2017 VISIR (dot-dashed line) and 2020 COMICS (dashed line) to indicate the expected range of the observations if the temperatures are unchanging since 2017. The location of local maxima (eastward) jets in the Cassini-derived wind field \citep{11garcia} are marked with vertical dotted lines.}
\label{postCassini}
\end{centering}
\end{figure}

Since the demise of Cassini in September 2017, ground-based facilities (including VLT/VISIR and Subaru/COMICS) have continued to track Saturn's mid-infrared emission.  VLT/VISIR was inaccessible in 2019-20, and Subaru/COMICS was decommissioned in 2020 (planned observations in 2019 failed).  Nevertheless, observations between 2017 and 2022 are available to test the evolution of the equatorial oscillation and the expectation from Fig. \ref{figure:11 radmodel} that the NPSV would continue to brighten until around 2021-22.  Brightness temperatures averaged within $\pm30^\circ$ of the central meridian are shown in Fig. \ref{postCassini} at 7.9, 12.3 and 17.7 $\mu$m.  These profiles were calibrated by scaling to Cassini/CIRS mid-latitude brightnesses in 2017 (from $15^\circ$N to $45^\circ$N), avoiding changes associated with the equatorial oscillation and polar vortices.  Latitudinal variability observed in these zonal averages is an order of magnitude larger to the standard error on the zonal mean. We also computed synthetic VISIR and COMICS images for 2017 and 2020, respectively, but both based on 2017 temperatures and hydrocarbon distributions - these are shown as the dotted lines in Fig. \ref{postCassini}, and reveal how mid-IR emission would change based solely on the small shift in viewing geometry over this time period, and the small change in plate scale between the COMICS and VISIR detectors.  At our mid-latitude reference point of $15-45^\circ$N, Fig. \ref{postCassini} shows that we would expect brightness changes of 1K, 2 K, 0.5 K at 7.8-$\mu$m, 12.2 $\mu$m and 17.6 $\mu$m (respectively), simply due to the change in geometry and instrument.  Unfortunately, this change cannot be captured by scaling solely to 2017 Cassini/CIRS measurements, so only variations exceeding these values can be considered as robust.

At 7.9 $\mu$m (Fig. \ref{postCassini}a), the brightness temperatures are remarkably consistent from 2017 to 2020, such that the 2020 COMICS observations can be largely reproduced with the 2017 CIRS measurements.  The final measurements in 2021 and 2022 then show a rather sudden shift in temperature, with the pole being $\sim5$ K warmer than the 2017-2020 observations.  The 2021 and 2022 VISIR observations both featured imaging on two nights, and the enhanced equator-to-pole gradient was seen in all four images, so is not an artefact of terrestrial observing conditions, nor were any background flux gradients identified across the detector.  The greater contrast between the NPSV and mid-latitudes can also be seen qualitatively in the 7.9 $\mu$m images in Fig. \ref{figure:2 thermal_images_comics_visir_pt1}.  Following the predictions of \citet{14guerlet} and \citet{16hue} in Fig. \ref{figure:11 radmodel}, these 2021-22 observations should be capturing the NPSV at its brightest, with a decline expected in subsequent years.  The 2021-22 brightness scans show a stronger equator-to-pole gradient than all images examined thus far.  

At 12.2 $\mu$m (Fig. \ref{postCassini}b), almost all of the observed brightness change (2-3 K) falls within the expected range based on a geometry shift alone.  The basic appearance of the 12.2-$\mu$m banded structure remains unchanged, but we see that both the 2020 COMICS and 2021-22 VISIR observations show a stronger equator-to-pole gradient than the 2017-18 data, again consistent with ongoing warming of the NPSV predicted by the radiative and photochemical models in Fig. \ref{figure:11 radmodel}. 

At low latitudes, the 2017 and 2018 7.9-$\mu$m data confirm a low-latitude temperature minimum, but the 2020-2022 observations reveal the redevelopment of the temperature maximum last seen in 2013.  This warm band still looks nothing like the strong temperature maximum observed at the last northern summer solstice \citep{89gezari, 08orton_qxo}.  Ethane emission at 12.4 $\mu$m shows a clearer local maximum at the equator in all years, indicating that the 7.9 and 12.4 $\mu$m filters sound different altitude regions of the vertical wave pattern.  

Finally, the belt/zone contrasts visible at 17.7 $\mu$m (Fig. \ref{postCassini}c) are very similar from 2017 to 2022, such that the connection between temperatures and winds (Section \ref{beltzone}) holds throughout this time period.  Only small changes are expected at 17.7 $\mu$m based on the changing geometry, but the latitudinal changes at the $\sim1$-K level are likely due to variable observing conditions between the four observing runs.  Nevertheless, the north polar cyclone (NPC) appears to be at its brightest in 2022 ($\sim97$ K) than at any previous point in this time series, and the surrounding polar troposphere (poleward of approximately $50^\circ$N) appears to have warmed by 3-4 K since 2016, likely in response to continued seasonal warming of the NPSV.

For both the low-latitude oscillations, and the evolution of the NPSV at high latitudes, the changes from 2017 to 2022 are often rather subtle.  Trends like the reestablishment of the local equatorial maximum, and the predicted cooling of the NPSV after 2022, will become clearer as the time series is extended into the 2020s.

%%%%%%%%%%%%%%%%%%%%%%%%%%%%%%%%%%%%%%%%%%%%%%
%%%%%%%%%%%%%%%%%%%%%%%%%%%%%%%%%%%%%%%%%%%%%%
%%%%%%%%%%%%%%%%%%%%%%%%%%%%%%%%%%%%%%%%%%%%%%

\section{Conclusion}
\label{summary}

Ground-based observatories have been monitoring Saturn's mid-infrared emission for almost four decades (1984-2022), spanning more than a full Saturnian year ($L_s=50^\circ$ to $L_s=140^\circ$) and providing seasonal context for observations by the Cassini spacecraft (2004-2017, $L_s=293-93^\circ$).  This study used filtered images between 7 and 25 $\mu$m to tackle three themes:  (i) the ability of Cassini-derived temperature and composition to reproduce the seasonal asymmetries and belt/zone structure observed in high-resolution images from 8-m primary mirrors; (ii) the repeatability of Saturn's temperature asymmetries, equatorial oscillations, and warm polar vortices from one Saturn year to the next; and (iii) the continued evolution of Saturn's northern summertime temperatures in the five years since the demise of Cassini.  The quality of infrared imaging has improved tremendously during the time span considered here, from the raster-scanned images from the 3-m IRTF (1984-1991), to 2D cameras (including MIRAC2 1993-1996 and MIRLIN 1996-2004), to the VLT/VISIR (2008-2022) and Subaru/COMICS (2005-2020) instruments on 8-m-class facilities.  This change in observational quality, combined with the difficulty of consistent radiometric calibrations, makes quantitative assessments of atmospheric variability rather challenging.  Nevertheless, this long-term survey allows us to draw the following conclusions:

\begin{enumerate}
    \item \textbf{Seasonal Contrasts: } Synthetic images generated using Cassini-derived temperatures and composition successfully reproduce seasonal trends and the banded structure observed in thermal images from VLT/VISIR and Subaru/COMICS, including seasonal brightness temperature changes of $\sim30$ K in the stratosphere and $\sim10-15$ K in the troposphere from solstice to solstice.  The latitudes and contrasts associated with Saturn's banded thermal structure are reproduced very well, as are the timescales for the growth and dissipation of emission from the polar stratospheric vortices (NPSV and SPSV) and compact polar cyclones (NPC and SPC) that were observed between 2004 and 2017.  The brightest north polar emission during the mission was observed in 2017, consistent with the continued warming of the NPSV at northern summer solstice ($L_s=90^\circ$, May 2017).  In most cases, this validates the Cassini record of temperatures and hydrocarbons, but we note that images near 10.7 $\mu$m (sensitive to PH$_3$) were not well reproduced, suggesting that latitudinal and temporal variability of PH$_3$ must be taken into account in future studies.  By stacking the filtered images to form a crude spectrum, we are able to estimate the vertical temperature structure as a function of latitude for the Earth-facing hemisphere, reproducing that observed by Cassini to within $\sim5$ K.
    
    \item \textbf{Belt/zone structure:} VISIR and COMICS images reveal the banded structure of Saturn's atmosphere, displaying 4-5 warm zonally-symmetric bands in each hemisphere.  These bands are present in filters sensing all altitudes from $\sim500$ mbar to the $\sim1$-mbar level.  Converting the latitudinal brightness temperature gradients into a `pseudo-shear' analogous to vertical windshear, we find a good negative correlation between the zonal wind speed and the strength of the shear.  This implies that westward jets experience eastward (positive) shear, and eastward jets experience westward (negative) shear throughout the troposphere and stratosphere, consistent with the decay of the zonal winds with altitude.  This strong connection between the winds and the banded thermal structure provides a better definition of Saturn's belts and zones than the muted albedo contrasts observed in visible light, and the fine-scale structure observed at 5 $\mu$m \citep{11fletcher_vims}.  This banded structure is suggestive of Ferrel-like circulation cells on the scale of Saturn's zonal jets, with rising motions in cool zones, and subsidence in warmer belts.  The same pattern is observed in Jupiter's upper troposphere \citep{20fletcher_beltzone}.  Thermal images from VLT and Subaru can monitor the banded contrasts (and windshear) associated with these circulation cells, and even reveal the hexagonal structure of the warm band near to the north pole.  The thermal hexagon is observed at the edge of the NPSV in both troposphere- and stratosphere-sensing filters, confirming that the hexagon extends hundreds of kilometers above the visible clouds.
    
    \item \textbf{Interannual Variability:} Observations from the IRTF allow us to study the thermal structure of Saturn over a full Saturnian year, and to compare to the predictions of radiative-climate and radiative-photochemical models.  Although systematic differences between observations and models persist, the models successfully reproduce the timing of polar enhancements in temperatures in each year, lagging behind the solstices by a phase difference that increases with pressure.  This good correspondence suggests that Saturn's thermal field is largely repeatable from year to year.  The first occurrence of the NPSV in 1984-1985 and 2014, and SPSV in 2000, are consistent with both models, confirming that this is primarily a radiative heating effect, albeit with a sharp boundary to the polar vortices introduced by dynamics. Furthermore, the tropospheric polar warming observed at 17-20 $\mu$m is also successfully reproduced by the models.   However, the temporal sampling of our existing dataset is insufficient to distinguish the subtle differences between models that hold hydrocarbons fixed \citep{14guerlet} and those that allow them to vary due to photochemistry \citep{16hue}.  The only sources of interannual variability, therefore, are dynamic phenomena that are not tied to Saturn's seasons, such as storm eruptions \citep{12fletcher} or the equatorial oscillation (see below).
    % The 1990s data hinted that the NPSV cooled faster than predicted by either model (possibly due to enhanced radiative cooling from aerosols), but this was not repeated in the 2010s and may have been attributable to differences in observational quality. 
    \item \textbf{Equatorial Oscillation: } Mid-IR images acquired one Saturnian year apart, by VISIR in 2018 and the observations of \citet{89gezari} in 1989, reveal changes in the equatorial stratospheric 7.9-$\mu$m emission that are not consistent with a $\sim15$-year `semi-annual' period of Saturn's Equatorial Stratospheric Oscillation \citep{08orton_qxo}. The warm equatorial band observed in 1989 was not present in 2018, although data in 2020-22 reveal the reemergence of the warm band.  A similar inter-annual variation between Voyager and Cassini observations one year apart had been noted previously \citep{14sinclair, 16fletcher}.  Qualitative inspection of the imaging data suggests either (i) a longer $\sim20$-year period for the 7.9-$\mu$m oscillation; or (ii) interruptions to the progression of the equatorial oscillation by tropospheric meteorology, leading to natural variability from year to year.  Intriguingly, infrared images in 1990 showed strong thermal perturbations in the equatorial stratosphere, possibly associated with the 1990 equatorial storm \citep{91sanchez} in the same way as was observed in 2011 \citep{12fletcher}.  Continued observations in the 2020s will be needed to re-establish the period and phase of the equatorial stratospheric oscillation.
    
    \item \textbf{Beyond Cassini:} Mid-infrared images from 2017-2022 have been used to extend the legacy of the Cassini observations, confirming continued atmospheric evolution through northern summer, approaching autumn equinox in May 2025 ($L_s=180^\circ$).  In particular, the NPSV continued to warm by approximately $\sim5$ K at the 0.5-5.0 mbar level, and 3-4 K at the 100-200 mbar level, by August 2022 ($L_s=148^\circ$), and the warm equatorial stratospheric band re-emerged in $\sim$2020 for the first time since 2013.  Radiative models predict that the NPSV should now begin to cool as it recedes from view, with the SPSV becoming visible again in the early 2030s.
    
\end{enumerate}

The long-term time series of infrared emission provides a valuable test of both the Cassini record of Saturn's temperatures, and the radiative climate models developed using different assumptions about photochemistry and dynamics.  Observations from 8-m-class facilities also reveal Saturn's banded circulation patterns and polar vortices in excellent detail, allowing us to monitor dynamic phenomena on the Earth-facing hemisphere.  However, ambiguous conclusions arise even when we have four decades of observations, as short-term dynamics can perturb the expected long-term seasonal and equatorial oscillations.  Short annual imaging runs offer an efficient means to continue to track Saturn's thermal variability, but additional progress could be made with high-resolution spectroscopy from the ground \citep[e.g., from TEXES,][]{02lacy} or space (e.g., JWST), providing stronger constraints on Saturn's vertical temperature structure and composition.

%%%%%%%%%%%%%%%%%%%%%%%%%%%%%%%%
%%%%%%%%%%%%%%%%%%%%%%%%%%%%%%%% ACKNOWLEDGEMENTS
%%%%%%%%%%%%%%%%%%%%%%%%%%%%%%%%
\section*{Acknowledgments}

Blake, Fletcher, Melin, Roman, Bardet and Antu\~{n}ano were  supported by a European Research Council Consolidator Grant (under the European Union's Horizon 2020 research and innovation programme, grant agreement No 723890) at the University of Leicester. Some of this research was carried out at the Jet Propulsion Laboratory, California Institute of Technology, under a contract with the National Aeronautics and Space Administration (80NM0018D0004).  This research used the ALICE High Performance Computing Facility at the University of Leicester.   We wish to thank the many observers and telescope operators who work long nights to provide the community with the data needed to advance planetary science, including (but not limited to) Dan Gezari, Thierry Fouchet, and Padma Yanamandra-Fisher.  We thank Sandrine Guerlet and Vincent Hue for making their radiative model predictions available to the community.  We thank student interns Charlotte Mason and Zheyu Li (Oxford University), supported by Caltech's Summer Undergraduate Research Fellowship program at JPL; and Shana Mastoon (Pasadena City College) supported by JPL's Student Independent Research Internship program.  Each of these students contributed to preliminary reductions of the datasets presented in this paper.  Finally, we thank three anonymous reviewers for their careful assessment of this manuscript.

The NEMESIS radiative transfer and spectral retrieval tool \citep{08irwin} is available via \url{https://github.com/nemesiscode/radtrancode}.  The Cassini/CIRS seasonal record of retrieved temperature and hydrocarbons is available here: \url{https://github.com/leighfletcher/CassiniCIRS}.

This study was based on observations collected at the European Organisation for Astronomical Research in the Southern Hemisphere; on data collected at the Subaru Telescope, which is operated by the National Astronomical Observatory of Japan; and on data collected from the Infrared Telescope Facility, which is operated by the University of Hawaii under contract 80HQTR19D0030 with the National Aeronautics and Space Administration.  Programme numbers are listed, where available, in Table \ref{datatable}.   The authors wish to recognise and acknowledge the very significant cultural role and reverence that the summit of Maunakea has always had within the indigenous Hawaiian community.  We are most fortunate to have the opportunity to conduct observations from this mountain.

For the purpose of open access, the author has applied a CC BY public copyright licence to any Author Accepted Manuscript version arising.

\appendix
\section{Data}

\begin{table*}
    \centering
    \scriptsize
    \begin{tabular}{|c|c|c|c|c|}
    \hline
      \textbf{Date} & \textbf{Telescope} & \textbf{Instrument} & \textbf{Prog. ID} \\  
    \hline
    
% Graber:
1984-05-01/05 & IRTF & Graber Rasters & - \\
1984-06-05/07 & IRTF & Graber Rasters & - \\
1985-05-18/21 & IRTF & Graber Rasters & - \\
1986-06-17/18 & IRTF & Graber Rasters & - \\
1987-06-07 & IRTF & Graber Rasters & - \\
1988-09-25 & IRTF & Graber Rasters & - \\
1989-04-23 & IRTF & Graber Rasters & - \\
1989-06-09 & IRTF & Graber Rasters & - \\
1989-08-28/29 & IRTF & Graber Rasters & - \\
1990-06-15 & IRTF & Graber Rasters & - \\
1990-08-24/27 & IRTF & Graber Rasters & - \\
\hline
% Gezari:
1989-03-28 & IRTF & Gezari Camera & - \\
1990-12-02/04 & IRTF & Gezari Camera & - \\
\hline
% MIRAC
1993-08-11/12 & IRTF & MIRAC2 & - \\
1995-11-05 & IRTF & MIRAC2 & - \\
1997-09-19/20 & IRTF & MIRAC2 & - \\
\hline
% SPECTROCAM-10
1995-06–19 & Palomar & SPECTROCAM-10 & - \\
1996-06-29 & Palomar & SPECTROCAM-10 & - \\ % (7.8 µm)
\hline
% MIRLIN:
1995-07-06 & IRTF & MIRLIN & - \\
1999-08-12 & IRTF & MIRLIN & - \\
2000-12-30 & IRTF & MIRLIN & - \\
2002-02-07 & IRTF & MIRLIN & - \\
2003-02-25 & IRTF & MIRLIN & - \\
%2004-02-09 & IRTF & MIRLIN & \\ %(not used)
\hline
% COMICS
% 2003-12-01 & Subaru & COMICS & \\
% 2004-11-04 & Subaru & COMICS & \\
2005-04-30 & Subaru & COMICS & o05108\\
2007-12-12 & Subaru & COMICS & o07173 \\
2008-01-23 & Subaru & COMICS & o07173\\
2009-01-13/14 & Subaru & COMICS & o08153\\
2013-05-01 & Subaru & COMICS & o13137\\
2017-05-17 & Subaru & COMICS & o17408\\
2017-09-06/07 & Subaru & COMICS & o17239\\
2018-05-25/26 & Subaru & COMICS & o18126\\
2018-08-20/21 & Subaru & COMICS & o18219\\
2019-05-26/27 & Subaru & COMICS & o19201\\
2020-06-03 & Subaru & COMICS & o20205\\
2020-07-31 & Subaru & COMICS & o20111\\
\hline
% VISIR
2008-04-12/15 & VLT & VISIR & 381.C-0560\\
2008-05-17/22 & VLT & VISIR & 381.C-0560\\
2008-06-10  & VLT & VISIR & 381.C-0560\\
2009-04-20/24   & VLT & VISIR & 383.C-0164 \\
2009-06-01/22/30    & VLT & VISIR & 383.C-0164\\
2009-12-24 & VLT & VISIR & 383.C-0164\\
2010-03-18/21 & VLT & VISIR & 084.C-0193\\
2010-04-01/23 & VLT & VISIR & 084.C-0193\\
2011-01-19/27/31 & VLT & VISIR & 386.C-0096\\
2011-02-08 & VLT & VISIR & 386.C-0096\\
2011-03-25/26 & VLT & VISIR & 386.C-0096\\
2011-05-25 & VLT & VISIR & 386.C-0096\\
2011-06-26 & VLT & VISIR & 287.C-5032\\
2011-07-20/24 & VLT & VISIR & 287.C-5032\\
2015-05-21 & VLT & VISIR & 095.C-0142\\
2015-08-27 & VLT & VISIR & 095.C-0142\\
2016-06-08 & VLT & VISIR & 097.C-0226\\
2016-07-29 & VLT & VISIR & 097.C-0226\\
2017-09-07/18 & VLT & VISIR & 099.C-0614\\
2018-05-29/30 & VLT & VISIR & 0101.C-0047\\
2018-07-16/28 & VLT & VISIR & 0101.C-0047\\
2021-11-01/03 & VLT & VISIR & 108.22B\\
2022-07-04 & VLT & VISIR & 109.2360\\
2022-08-14 & VLT & VISIR & 109.2360\\
 \hline   

    \end{tabular}
    \caption{Mid-IR imaging data used in this study.  Where dates are given with a range, the best observations from a multi-day observing run were selected.}
    \label{datatable}
\end{table*}

\bibliographystyle{elsarticle-harv}
\bibliography{references}




%% \bibitem must have one of the following forms:
%%   \bibitem[Jones et al.(1990)]{key}...
%%   \bibitem[Jones et al.(1990)Jones, Baker, and Williams]{key}...
%%   \bibitem[Jones et al., 1990]{key}...
%%   \bibitem[\protect\citeauthoryear{Jones, Baker, and Williams}{Jones
%%       et al.}{1990}]{key}...
%%   \bibitem[\protect\citeauthoryear{Jones et al.}{1990}]{key}...
%%   \bibitem[\protect\astroncite{Jones et al.}{1990}]{key}...
%%   \bibitem[\protect\citename{Jones et al., }1990]{key}...
%%   \harvarditem[Jones et al.]{Jones, Baker, and Williams}{1990}{key}...
%%


% \end{thebibliography}

\end{document}